\documentclass[12pt,a4paper]{article}
\usepackage{graphicx}
\usepackage{amssymb, amsmath}

\usepackage{amsthm}
\usepackage{fullpage}

\makeatletter

\theoremstyle{definition}
\newtheorem{definition}{Definition}
\newtheorem*{remark*}{Remark}

\theoremstyle{plain}
\newtheorem{theorem}[definition]{Theorem}
\newtheorem{lemma}[definition]{Lemma}
\newtheorem*{theorem*}{Theorem}

\def\epsilon{\varepsilon}

\def\myscaling{0.72}   

\def\myurl{\texttt{
http://www.csc.liv.ac.uk/\nolinebreak[3]$\sim$markus/\nolinebreak[3]kagome5colours/}}

\newcommand\vbp{\mathcal{X}}
\newcommand{\exR}{\mathcal{R}}

\newcommand{\dtv}{\mathrm{d}_\mathrm{TV}}
\newcommand{\ball}{\mathrm{Ball}}

\newcommand{\calA}{\mathcal{A}}
\newcommand{\calF}{\mathcal{F}}
\newcommand{\calS}{\mathcal{S}}

\makeatother

\begin{document}

\title{Strong spatial mixing and rapid mixing with five colours for the kagome lattice}

\author{Markus Jalsenius\\~\\Department of Computer Science\\University of Liverpool\footnote{Some of the work has been done at the Department of Computer Science,
University of Warwick, Coventry, CV4~7AL, UK.}\\Liverpool, L69 3BX, UK}


\maketitle

\begin{abstract}
We consider proper 5-colourings of the kagome lattice. Proper $q$-colourings
correspond to configurations in the zero-temperature $q$-state anti-ferromagnetic
Potts model. Salas and Sokal have given a computer assisted proof
of strong spatial mixing on the kagome lattice for $q\geq6$ under
any temperature, including zero temperature. It is believed that there
is strong spatial mixing for $q\geq4$. Here we give a computer assisted proof of strong
spatial
mixing for $q=5$ and zero temperature. It is commonly known that strong spatial mixing
implies that
there is a unique infinite-volume Gibbs measure and that the Glauber
dynamics is rapidly mixing. We give a proof of rapid mixing
of the Glauber dynamics on any finite subset of the vertices of the kagome lattice,
provided that the boundary is free (not coloured). The Glauber dynamics is not necessarily
 irreducible if the boundary is chosen arbitrarily for $q=5$ colours.
The Glauber dynamics can be used to uniformly sample proper 5-colourings.
Thus, a consequence of rapidly mixing Glauber dynamics is that there
is fully polynomial randomised approximation scheme for counting the
number of proper 5-colourings.
\end{abstract}

\section{Introduction}

Proper colourings correspond to configurations in the \emph{zero-temperature
anti-ferromagnetic Potts model}. In this paper we will show that the
system specified by proper 5-colourings of the \emph{kagome lattice}
has \emph{strong spatial mixing}, and that the \emph{Glauber dynamics}
is \emph{rapidly mixing}. The previously best known result~\cite{ss-apt-97}
on mixing on the kagome lattice was for 6 colours. It is believed~\cite{ss-apt-97}
that there is strong spatial mixing for 4 or more colours, and
hence our result is narrowing the gap between what is believed and known.
In Section~\ref{sub:definitions-and-background} below we give an
introduction to mixing, and in Section~\ref{sec:results} we state
our results and discuss related work.

\subsection{Definitions and background}

\label{sub:definitions-and-background}
\begin{figure}[tp]
\centering
(a)\hspace{-6mm}\includegraphics[scale=\myscaling]{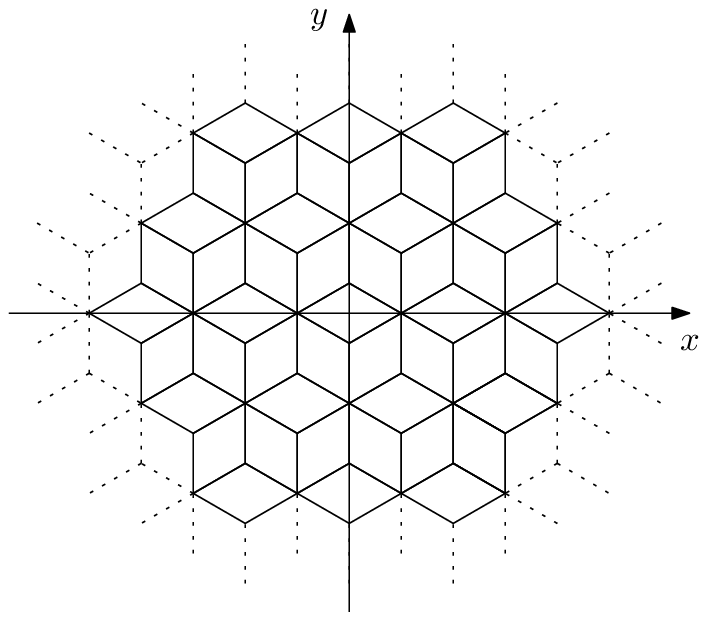}
(b)\hspace{-2mm}\includegraphics[scale=\myscaling]{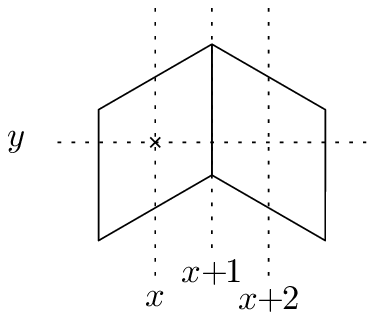}\hspace{6mm}
(c)\hspace{-2mm}\includegraphics[scale=\myscaling]{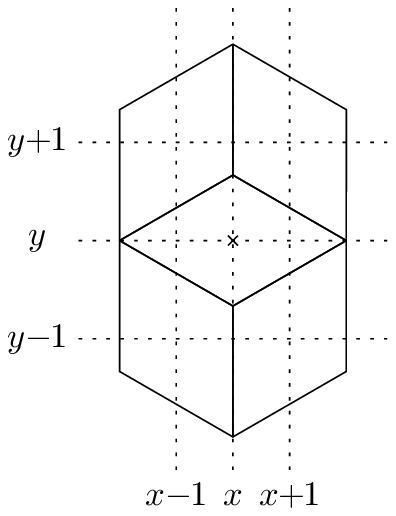}

\smallskip

\caption{\label{fig:kagome_lattice}(a) The kagome lattice, here drawn in a
coordinate system. We illustrate graphs such that a face represents
a vertex. (b) A vertex $(x,y)\in V_{\textrm{odd}}$ and its right
neighbour. (c) A vertex $(x,y)\in V_{\textrm{even}}$ and its four
neighbours.}
\end{figure}

The kagome lattice,
Figure~\ref{fig:kagome_lattice}(a), is a natural lattice of interest in statistical
physics~\cite{ss-apt-97}.
Instead of drawing graphs in the traditional way, with a vertex denoted
with a solid circle and an edge denoted with a line segment, we
draw graphs such that faces represent vertices. Two adjacent faces
therefore represent two adjacent vertices.
Let $\mathcal{G} = (V_\mathcal{G}, E_\mathcal{G})$
denote the kagome lattice
with vertex set $V_\mathcal{G}$ and edge set $E_\mathcal{G}$.
We have $V_{\mathcal{G}}=V_{\textrm{odd}}\cup V_{\textrm{even}}$, where
\begin{eqnarray*}
V_{\textrm{odd}} & = & \{(x,y)\;\mid\; x,y\in\mathbb{Z}\textup{ are both odd}\},\\
V_{\textrm{even}} & = & \{(x,y)\;\mid\; x=4k_{1}+r,\; y=4k_{2}+r,\;
k_{1},k_{2}\in\mathbb{Z},\; r\in\{0,2\}\}.
\end{eqnarray*}
The edge set
\begin{eqnarray*}
E_{\mathcal{G}} & = & \{((x,y),(x+2,y))\;\mid\;(x,y)\in V_{\textrm{odd}}\}\;\cup\\
&  & \{((x,y),(x+k_{1},y+k_{2}))\;\mid\;(x,y)\in V_{\textrm{even}},\; k_{1}\in\{-1,1\},\;
k_{2}\in\{-1,1\}\}.
\end{eqnarray*}
Note that both vertices in Figure~\ref{fig:kagome_lattice}(b) are in $V_{\textrm{odd}}$,
and we see that two adjacent vertices in $V_{\textrm{odd}}$ differ by 2 in their
$x$-coordinate. In Figure~\ref{fig:kagome_lattice}(c), the centre vertex is in
$V_{\textrm{even}}$, and we see that its four neighbours are in $V_{\textrm{odd}}$.

A \emph{region} $R\subseteq V_{\mathcal{G}}$ is a \emph{finite} \emph{non-empty}
subset of the vertex set of the kagome lattice. The subset
$\partial R\subseteq V_{\mathcal{G}}$
denotes the \emph{vertex boundary} of $R$ such that $\partial R$
is the set of vertices that are not in $R$ but are adjacent to any
vertex in $R$. The edge set $E(R)$ is the set of all edges
$(u,v)\in E_{\mathcal{G}}$ such that at least one of the vertices
$u$ and $v$ is in $R$. The \emph{edge boundary} $\mathcal{E}R$
of $R$ is the set of all edges $(u,v)\in E(R)$ such that
exactly one of the vertices $u$ and $v$ is in $R$ and the other
one is in $\partial R$.

The set $Q=\{1,\dots,q\}$ denotes the set of $q$ \emph{colours},
and the set $Q_{0}=\{0\}\cup Q$. The colour~0 represents {}``no colour''.
A $q$-\emph{colouring} of a region $R$ is a function from $R$ to
the set $Q$, and a $q_{0}$-colouring of $R$ is a function from
$R$ to $Q_{0}$. A 0-colouring of $R$ is a function from $R$ to
the set $\{0\}$, which means that all vertices in $R$ are assigned
colour~0. We often write only colouring when it is obvious from the
context if it is a $q$-, $q_{0}$- or 0-colouring, or if any colouring
will do. Let $\sigma$ be a colouring of a region $R$. If $R'$ is
a subset of $R$ then $\sigma(R')$ is the colouring of $R'$ induced
by $\sigma$. Furthermore,
for a vertex $v\in R$,
$\sigma(v)$ is the colour of $v$
under $\sigma$. Let $\Omega_{R}^{+}$ denote the set of all $q$-colourings
of the region $R$. For two colourings $\sigma,\sigma'\in\Omega_{R}^{+}$,
the \emph{Hamming distance} between $\sigma$ and $\sigma'$ is the
number of vertices in $R$ on which $\sigma$ and $\sigma'$ differ.
A colouring $\sigma$ of $R$ is \emph{proper} if no adjacent vertices
receive the same colour. That is, $\sigma(u)\neq\sigma(v)$ for all
adjacent vertices $u$ and $v$ in $R$. Let $\Omega_{R}$ denote
the set of all proper $q$-colourings of the region $R$. Given a
$q_{0}$-colouring $\mathcal{B}$ of $\partial R$, a proper $q$-colouring
$\sigma$ of $R$ \emph{agrees} with $\mathcal{B}$ if $\sigma(u)\neq\mathcal{B}(v)$
for all $(u,v)\in\mathcal{E}R,$ where $u\in R$. We let $\Omega_{R}(\mathcal{B})$
denote the set of all proper $q$-colourings of $R$ that agree with
$\mathcal{B}$. The uniform distribution on $\Omega_{R}(\mathcal{B})$
is denoted $\pi_{\mathcal{B}}$, and for any subregion $R'\subseteq R$,
let $\pi_{\mathcal{B},R'}$ denote the distribution on proper $q$-colourings
of $R'$ induced by $\pi_{\mathcal{B}}$. 

In this paper we will show that the system specified by proper 5-colourings
of the kagome lattice has \emph{strong spatial mixing}. Informally, strong
spatial mixing means that if $R$ is a region and $\mathcal{B}$ is
a $q_{0}$-colouring of $\partial R$, then the effect the colour
of a vertex $w\in\partial R$ has on a vertex $v\in R$ decays exponentially
with the distance between $w$ and $v$. The effect is measured with
the \emph{total variation distance}. For two distribution $D_{1}$
and $D_{2}$ on a set $S$, the total variation distance between $D_{1}$
and $D_{2}$ is defined as\[
\dtv(D_{1},D_{2})=\frac{1}{2}\sum_{s\in S}|D_{1}(s)-D_{2}(s)|=
\max_{A\subseteq S}|D_{1}(A)-D_{2}(A)|.\]
The following definition of strong spatial mixing is taken from~\cite{gmp-ssmfc-04}
and is adapted to the kagome lattice.

\begin{definition}
[Strong spatial mixing]
\label{def:strong-spatial-mixing}
The system
specified by proper $q$-colourings of the kagome lattice has \emph{strong
spatial mixing} if there are two constants $\alpha>0$ and $\epsilon \in (0,1)$
such that, for any region~$R$, any subregion $R'\subseteq R$,
any two $q_{0}$-colourings $\mathcal{B}$ and $\mathcal{B}'$
of $\partial R$ which differ on exactly one vertex $w\in\partial R$
and such that 
$\mathcal{B}(w)\neq0$ and $\mathcal{B}'(w)\neq0$,
$$
\dtv(\pi_{\mathcal{B},R'},\pi_{\mathcal{B}',R'})\leq\alpha|R'|(1-\epsilon)^{d(w,R')},
$$
where $d(w,R')$ is the minimal distance within~$R$ from $w$ to some
vertex of $R'$.
\end{definition}

A distribution $\pi$ on the set of proper $q$-colourings of the
infinite kagome lattice is an \emph{infinite-volume Gibbs distribution} if,
for any region $R$ and any proper $q$-colouring $\sigma$ of the
kagome lattice, the conditional distribution
$\pi(\cdot|\sigma(V_{\mathcal{G}}\backslash R))$
on $\Omega_{R}$ (conditioned on the colouring $\sigma(V_{\mathcal{G}}\backslash R)$
of all vertices other than those in $R$) is $\pi_{\mathcal{B}}$,
where $\mathcal{B}=\sigma(\partial R)$. It is known that there is
always at least one infinite-volume Gibbs distribution, and the question
of interest is to determine whether it is unique or not. This question
is central in statistical physics because it corresponds to the number
of macroscopic equilibria for a given system. The phenomenon of non-uniqueness
corresponds to what is referred to as a \emph{phase transition}.
A consequence of strong spatial mixing is that the infinite-volume
Gibbs distribution is unique \cite{dsvw-mtslss-04,w-mtsdss-04,w-ccugm-05}.
For
more on Gibbs distributions, see for example~\cite{g-gmpt-88} or~\cite{ghm-rgef-99}.

Another question of interest is to determine how quickly the system
converges to equilibrium.
The answer to this question is connected to the quantities $\alpha$ and $\epsilon$
in Definition~\ref{def:strong-spatial-mixing} above.
From a statistical physics point of view,
this question is important for understanding phenomena such as how
the system returns to equilibrium after a shock forces it out of it.
In this paper we consider a famous dynamical process called the
\emph{Glauber dynamics} which models how the system converges. The Glauber dynamics,
defined next, is a Markov chain that performs \emph{single-vertex heat-bath}
updates.

\begin{definition}
[Glauber dynamics]\label{def:glauber}For any region $R$ and any
$q_{0}$-colouring $\mathcal{B}$ of $\partial R$, the \emph{Glauber
dynamics} is a Markov chain with state space $\Omega_{R}(\mathcal{B})$,
and a transition is made from a state $\sigma$ to $\sigma'$ in the
following way:
\begin{enumerate}
\item Choose a vertex $v$ uniformly at random from $R$.
\item Let $Q_{v}$ be the set of colours which are assigned to the neighbours
of $v$ (either in $\sigma$ or $\mathcal{B}$).
\item Choose a colour $c$ uniformly at random from $Q\backslash Q_{v}$
and obtain the new colouring $\sigma'$ from $\sigma$ by assigning
colour $c$ to vertex $v$.
\end{enumerate}
\end{definition}

\begin{figure}[tp]
\centering
(a)\includegraphics[scale=\myscaling]{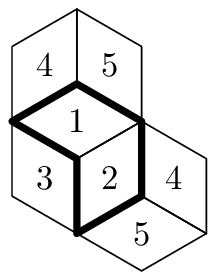}
\hspace{6mm}\includegraphics[scale=\myscaling]{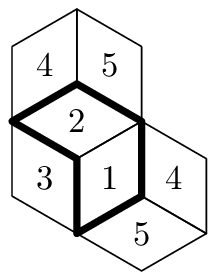}
\hspace{12mm}(b)\includegraphics[scale=\myscaling]{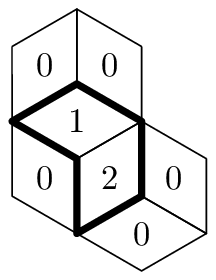}
\hspace{6mm}\includegraphics[scale=\myscaling]{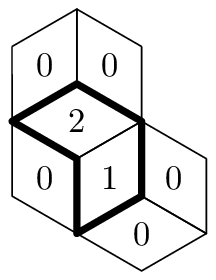}

\smallskip

\caption{\label{fig:frozen-states}A 2-vertex region of the kagome lattice
with different colourings. A vertex is labelled with its colour. The
two colourings in (a) are {}``frozen'' and do not communicate in
the Glauber dynamics for $q=5$. However, when restricting the boundary
to the 0-colouring in (b), the two colourings do communicate.}

\end{figure}

A sufficient condition for the Glauber dynamics to be \emph{connected}
(that is, any proper colouring can be obtain from another proper colouring
by a series of transitions) is to have $q\geq6$. In general, with
the Glauber dynamics defined similarly on any underlying infinite graph
of maximum degree $\Delta$,
having
$q\geq\Delta+2$ is a sufficient condition for
the dynamics to be connected.
In this paper
we focus on 5-colourings and in order to guarantee that the Glauber
dynamics is connected we will have to restrict the colourings $\mathcal{B}$
of the boundary to the 0-colouring (see Figure~\ref{fig:frozen-states}).%
For this reason, our 5-colour mixing result
for the Glauber dynamics is restricted to the 0-colouring of the boundary.
It is worth pointing out that if we add moves to the Glauber dynamics that allow swapping
the colours of two neighbouring vertices
(when this move is allowed with respect to the colouring of the rest of the vertices)
then the new dynamics is connected for any region $R$ and any $q_0$-colouring of
$\partial R$ if $q \geq 5$.
This fact is true for any graph of maximum degree $\Delta$ and
$q \geq \Delta + 1$ colours.
This augmented Glauber dynamics can be simulated by the \emph{heat-bath dynamics on edges}
which we define as follows: Choose an edge $e = (v_1, v_2)$ uniformly at random and
simultaneously recolour $v_1$ and $v_2$ uniformly at random from the allowed colourings.

If the Glauber dynamics is connected, and hence ergodic,
then $\pi_{\mathcal{B}}$ is the unique stationary distribution.
This follows from the fact that the Glauber dynamics is reversible
with respect to $\pi_{\mathcal{B}}$.
For the same reason, $\pi_{\mathcal{B}}$ is the unique stationary distribution
of the heat-bath dynamics on edges.
The Glauber dynamics can be used as a sampler to sample colourings from the uniform
distribution on $\Omega_{R}(\mathcal{B})$. This can be done efficiently
if the Glauber dynamics is \emph{rapidly mixing}
(see definition below), which means that it quickly
reaches its stationary distribution.

\begin{definition}
[Mixing time]
Consider the Glauber dynamics on a region $R$ with boundary colouring $\mathcal{B}$.
Let $P^{t}(\sigma,\sigma')$
be the probability of going from state $\sigma$ to $\sigma'$ in
exactly $t$ steps. For any $\delta>0$, the \emph{mixing time} \[
\tau(\delta)=\max_{\sigma\in\Omega_{R}(\mathcal{B})}\,\min\{t_{0}\,:\,
\dtv(P^{t}(\sigma,\cdot),\pi_{\mathcal{B}})\leq\delta\textup{ for all }t\geq t_{0}\}.\]
The Glauber dynamics is \emph{rapidly mixing} if $\tau(\delta)$ is upper-bounded
by a polynomial in the region-size $|R|$ and $\log(1/\delta)$.
\end{definition}

It is a well-known fact that if
the system has strong spatial mixing then the Glauber dynamics is
(often) rapidly mixing \cite{dsvw-mtslss-04,m-lgddsm-97,w-mtsdss-04}.
In Section~\ref{sec:rapid-mixing} we will study this fact and see
how strong spatial mixing and rapid mixing are closely related. For
$q\geq6$ colours (or $q\geq\Delta+2$ in general) it is straightforward
to apply Theorem~8 in~\cite{gmp-ssmfc-04} in order to infer rapid
mixing from strong spatial mixing. However, with $q=5$ colours we
cannot rely entirely on previous results. We will establish certain
properties of 5-colourings of the kagome lattice and show that the
Glauber dynamics is rapidly mixing for $q=5$ under the 0-colouring
of the boundary.

In~\cite{j-csi-03} it is explained how \emph{approximate counting}
and almost uniform sampling are related. If there is a method for
sampling (almost) uniformly at random in polynomial time from the
set of proper colouring of a finite region $R$, then we can construct
a \emph{fully polynomial randomised approximation scheme}, or \emph{FPRAS},
for counting the number of proper colourings of $R$. Thus, if the
Glauber dynamics is rapidly mixing then we could use it to construct (in a non-trivial
way) an FPRAS for estimating $|\Omega_{R}|$. For details
on the topic of how sampling and counting are related, see Jerrum~\cite{j-csi-03}
and Jerrum, Valiant and Vazirani~\cite{jvv-rgcs-86}.

\section{The results and related work}
\label{sec:results}

We will prove the following theorems, which
improve previously known results on mixing for proper colourings
of the kagome lattice.

\begin{theorem}
\label{thm:kagome-ssm}The system specified by proper 5-colourings
of the kagome lattice has strong spatial mixing.
\end{theorem}

\begin{theorem}
\label{thm:kagome-glauber}For any region $R$ of the kagome lattice
and $q=5$ colours, the Glauber dynamics is rapidly mixing on $R$
under the 0-colouring of $\partial R$. The mixing time
$\tau(\delta)\in O(n^{2}+n\log\frac{1}{\delta})$,
where $n$ is the number of vertices in $R$.
\end{theorem}

\begin{theorem}
\label{thm:edge-dynamics}For any region $R$ of the kagome lattice
and $q=5$ colours, the heat-bath dynamics on edges is rapidly mixing on $R$
under any $q_0$-colouring of $\partial R$. The mixing time
$\tau(\delta)\in O(n^{2}+n\log\frac{1}{\delta})$,
where $n$ is the number of vertices in $R$.
\end{theorem}

The previously best known result on mixing on the kagome lattice is
that of Salas and Sokal~\cite{ss-apt-97}. They provided a computer
assisted proof of strong spatial mixing for
$q = 6$ colours. It is believed~\cite{ss-apt-97} that there
is strong spatial mixing for $q\geq4$ colours.

It is worth mentioning some previous general results on mixing.
Independently, Jerrum~\cite{j-vsa-95} and Salas and Sokal~\cite{ss-apt-97}
proved that for proper $q$-colourings on a graph of maximum degree $\Delta$
the Glauber dynamics has $O(n \log n)$-mixing for $q > 2 \Delta$,
where $n$ is the number of vertices of the region.
For $q = 2 \Delta$, Bubley and Dyer~\cite{bd-pctprmmc-97} showed that
it mixes in $O(n^3)$ time and Molloy~\cite{m-vrmmc2dc-01} showed
that it mixes $O(n \log n)$ time.
Vigoda~\cite{v-ibsc-00} used a Markov chain that differs from the
Glauber dynamics and showed that it has $O(n \log n)$-mixing for
$q > (11/6)\Delta$. This result implied that the Glauber dynamics
is rapidly mixing for $q > (11/6)\Delta$.
Goldberg, Martin and Paterson~\cite{gmp-ssmfc-04}
showed that any triangle-free graph has
strong spatial mixing provided $q>\alpha\Delta-\gamma$, where
$\alpha$ is the solution to $\alpha^{\alpha}=e$ ($\alpha\approx1.76322$)
and $\gamma=\frac{4\alpha^{3}-6\alpha^{2}-3\alpha+4}{2(\alpha^{2}-1)}\approx0.47031$.
Note that their result cannot be applied to the kagome lattice since
its edge set contains triangles. However, for other 4-regular graphs, such as the square
lattice $\mathbb{Z}^{2}$, it follows that mixing occurs for $q\geq7$ colours.
The technique Goldberg, Martin and Paterson used in~\cite{gmp-ssmfc-04}
is well suited to be extended to involve special cases that depend on the
particular graph under consideration.
Involving such special cases can improve the mixing bounds.
In order to deal with all special cases it might be
helpful to incorporate computer assistance.
This has been done in~\cite{gmp-ssmfc-04} for the lattice
$\mathbb{Z}^{3}$. The general result gives mixing for $q\geq11$
colours but by taking advantage of the geometry of the lattice it has been
shown that mixing occurs for $q\geq10$. This proof is computer assisted.
Another computer assisted proof of mixing in~\cite{gmp-ssmfc-04}
is given for the triangular lattice and $q=10$ colours. This result
was improved by Jalsenius~\cite{j-ssmrm9ctl-07} to $q=9$ by exploiting the geometry of the
lattice even further.
Goldberg, Jalsenius, Martin and Paterson used the
technique from~\cite{gmp-ssmfc-04} and gave in~\cite{gjmp-imbafpm-06} a
computer assisted proof of mixing for $q=6$ on the square lattice
$\mathbb{Z}^{2}$. This is an alternative proof of the result of Achlioptas,
Molloy, Moore and van Bussel~\cite{ammb-sgcfc-04} (who also
used computer assistance).
In this paper we will refine the
technique Goldberg, Martin and Paterson introduced in~\cite{gmp-ssmfc-04}
to show mixing on the kagome lattice for $q=5$ colours. Both the square lattice and the
kagome lattice are 4-regular graphs, but the kagome lattice contains triangles
whereas the square lattice does not.
An interesting observation is that the presence of triangles seem to have
a positive effect on the technique we use to show strong spatial mixing.
Attempts to prove mixing with 5 colours on the square lattice with this
technique has failed so far. The absence of triangles seem to be one strong
reason why
(assuming the square lattice does have strong spatial mixing with 5~colours).

\section{The framework}

When Goldberg, Martin and Paterson~\cite{gmp-ssmfc-04} derived improved
mixing bounds for spin systems consisting of proper colourings, they
introduced the notion of a \emph{vertex-boundary pair}. A vertex-boundary
pair is a data structure holding information about a region $R$ and
colourings of $\partial R$. The idea is to derive certain
properties of the vertex-boundary pairs which can be easily translated
into properties such as whether there is strong spatial mixing or
not. When Goldberg, Martin and Paterson derived these properties,
it turned out to be convenient to work with \emph{edge-boundary pairs}.
An edge-boundary pair (defined in the next section) contains colourings
of the edge boundary $\mathcal{E}R$ rather than the vertex boundary
$\partial R$.

\begin{definition}
[Vertex-boundary pair]A \emph{vertex-boundary pair} $\vbp$ consists
of
\begin{itemize}
\item a region $R_{\vbp}$,
\item a distinguished boundary vertex $w_{\vbp}\in\partial R_{\vbp}$, and
\item a pair $(\mathcal{B}_{\vbp},\mathcal{B}_{\vbp}')$ of $q_{0}$-colourings
of $\partial R_{\vbp}$ that are identical on all vertices except
on $w_{\vbp}$, where they differ. The colour of $w_{\vbp}$ is in
$Q$ for both $\mathcal{B}_{\vbp}$ and $\mathcal{B}_{\vbp}'$.
\end{itemize}
\end{definition}
Note that the colour of the distinguished vertex $w_{\vbp}$ has to
be in the set $Q$. That is, $\mathcal{B}_{\vbp}(w_{\vbp})\neq0$
and $\mathcal{B}_{\vbp}'(w_{\vbp})\neq0$. Definition~\ref{def:strong-spatial-mixing}
of strong spatial mixing can be rephrased using the definition of
a vertex-boundary pair. That is, in order to show strong spatial mixing,
we will show that there are two constants $\alpha>0$ and $\epsilon \in (0,1)$
such that for every vertex-boundary pair $\vbp$ and every subregion
$R'\subseteq R_{\vbp}$, \[
\dtv(\pi_{\mathcal{B}_{\vbp},R'},\pi_{\mathcal{B}_{\vbp}',R'})\leq
\alpha|R'|(1-\epsilon)^{d(w_{\vbp},R')}.\]
One approach to show exponential decay of the total variation distance
in the distance between $w_{\vbp}$ and $R'$ is to construct a suitable
\emph{coupling} (defined next) of the distributions $\pi_{\mathcal{B}_{\vbp}}$ and
$\pi_{\mathcal{B}_{\vbp}'}$. For two distributions $D_{1}$ and $D_{2}$
on a set $S$, a coupling $\Psi$ of $D_{1}$ and $D_{2}$ is a joint
distribution on $S\times S$ with marginal distributions $D_{1}$
and $D_{2}$. If the pair $(X_{1},X_{2})$ is a random variable drawn
from $\Psi$ then \[
\dtv(D_{1},D_{2})\leq\Pr[X_{1}\neq X_{2}].\]
Thus, in order to upper-bound the total variation distance, one can
find some suitable coupling $\Psi$ and compute the probability of
having $X_{1}\neq X_{2}$. The aim here is to construct a coupling
$\Psi_{\vbp}$ of $\pi_{\mathcal{B}_{\vbp}}$ and $\pi_{\mathcal{B}_{\vbp}'}$
such that if the pair $(\sigma,\sigma')$ of colourings is drawn from
$\Psi_{\vbp}$ then the probability that $\sigma$ and $\sigma'$
differ on $R'\subseteq R_{\vbp}$ decreases exponentially with the
distance between the discrepancy vertex $w_{\vbp}\in\partial R_{\vbp}$
and $R'$. For a vertex $v\in R_{\vbp}$ we define the indicator random
variable $1_{\Psi_{\vbp},v}$ for the event that the colour of $v$
differs in a pair of colourings drawn from $\Psi_{\vbp}$. Hence,
the quantity $\sum_{v\in R_{\vbp}}\mathbb{E}[1_{\Psi_{\vbp},v}]$
is the expected number of vertices in $R_{\vbp}$ on which the colours
differ in a pair of colourings drawn from $\Psi_{\vbp}$.
If $\mathbb{E}[1_{\Psi_{\vbp},v}]$
is small enough for all vertex-boundary pairs $\vbp$ and vertices
$v\in R_{\vbp}$ then we can infer strong spatial mixing
(Section~\ref{sec:exp-decay-vertices-and-ssm})
and rapid mixing (Section~\ref{sec:rapid-mixing}).

\section{Edge discrepancies}

\label{sec:edge-discrepancies}Similarly to the definition of a vertex
colouring we define a $q$-, $q_{0}$- and $0$-colouring of a set
$E\subseteq E_{\mathcal{G}}$ of edges to be a function from $E$
to $Q$, $Q_{0}$ and $\{0\}$, respectively. If $B$ is an edge colouring
of $E$, and $E'$ is a subset of $E$ then $B(E')$ is the colouring
of $E'$ induced by $B$.
For an edge $e\in E$, $B(e)$ is the colour of $e$
under $B$. Given a region $R$ and a $q_{0}$-colouring
$B$ of $\mathcal{E}R$, a proper $q$-colouring $\sigma$ of $R$
agrees with $B$ if $\sigma(u)\neq B(e)$ for all edges $e\in\mathcal{E}R,$
where $u\in R$ is incident to $e$. We let $\Omega_{R}(B)$ denote
the set of all proper $q$-colourings of $R$ that agree with $B$.
The uniform distribution on $\Omega_{R}(B)$ is denoted $\pi_{B}$.

Let $E$ be a set that contains the four edges that are incident to
some vertex $v\in V_{\mathcal{G}}$. Two edges $e,e'\in E$ are \emph{adjacent}
if there is a clockwise ordering around $v$ of the edges in $E$
such that $e'$ follows immediately after $e$. Similarly to a vertex-boundary
pair $\vbp$ we define an edge-boundary pair $X$ as follows. Note
that this definition is equivalent to the notion of a relevant boundary-pair
in~\cite{gmp-ssmfc-04}.

\begin{definition}
[Edge-boundary pair]\label{def:edge-boundary-pair}An \emph{edge-boundary
pair} $X$ consists of
\begin{itemize}
\item a region $R_{X}$,
\item a distinguished boundary edge $e_{X}=(w_{X},v_{X})\in\mathcal{E}R_{X}$
with $w_{X}\in\partial R_{X}$, $v_{X}\in R_{X}$, and
\item a pair $(B_{X},B'_{X})$ of $q_{0}$-colourings of $\mathcal{E}R_{X}$
that are identical on all edges except on $e_{X}$, where they differ.
\end{itemize}
We require
\begin{itemize}
\item $B_{X}(e_{X})\in Q$ and $B_{X}'(e_{X})\in Q$, and
\item any two adjacent boundary edges that share a vertex in $\partial R_{X}$
have the same colour in at least one of the two colourings $B_{X}$
and $B'_{X}$ (and so in both of $B_{X}$ and $B'_{X}$ except when
edge $e_{X}$ is involved).
\end{itemize}
\end{definition}

Suppose $X$ is an edge-boundary pair.
For a coupling $\Psi_{X}$ of $\pi_{B_{X}}$
and $\pi_{B_{X}'}$ we define $1_{\Psi_{X},v}$ to be the indicator
random variable for the event that, when a pair of colourings is drawn
from $\Psi_{X}$, the colour of vertex $v\in R_{X}$ differs in these
two colourings. For any edge-boundary pair $X$ we define $\Psi_{X}^{\textup{min}}$
to be some coupling of $\pi_{B_{X}}$ and $\pi_{B_{X}'}$ minimising
$\mathbb{E}[1_{\Psi_{X},v_{X}}]$.
For every
pair of colours $c,c'\in Q$, let $p_{X}^{\textup{min}}(c,c')$ be
the probability
that $C(v_{X})=c$ and $C'(v_{X})=c'$,
where $(C,C')$ is a pair of colourings drawn from $\Psi_{X}^{\textup{min}}$.
For a vertex $v\in R_X$,
let $d(e_{X},v)$ denote the distance within
$R_{X}$ from edge $e_{X}$ to $v$. Thus, $d(e_{X},v_{X})=1$ and
if $v\in R_{X}$ adjoins $v_{X}$ then $d(e_{X},v)=2$, and so on.
We wish to construct a coupling $\Psi_{X}$ of $\pi_{B_{X}}$ and
$\pi_{B_{X}'}$ such that $\mathbb{E}[1_{\Psi_{X},v}]$ decreases
exponentially in the distance $d(e_{X},v)$. In order to do this
we use a recursive coupling. To aid the analysis we define a labelled
tree $T_{X}$ associated with each edge-boundary pair $X$. The notion
of $T_{X}$ was introduced by Goldberg, Martin and Paterson in~\cite{gmp-ssmfc-04}.

Suppose $X$ is an edge-boundary pair. We will now construct the tree $T_X$.
Start with a node $r$ which
will be the root of $T_{X}$. For every pair $c, c'\in Q$
of distinct colours,
add an edge labelled $(p_{X}^{\textup{min}}(c,c'),v_{X})$
from $r$ to a new node $r_{c,c'}$.
Let $e_1,e_2,e_3$ be the clockwise
ordering of the edges incident to $v_{X}$ (excluding edge $e_X$)
such that $e_{X}$ appears between $e_3$ and $e_1$.
The $i$-th neighbour of $v_X$ denotes the vertex that is incident to $e_i$.
If the $i$-th neighbour of $v_X$ is not in $R_X$ then
we define $X_{i}(c,c') = \emptyset$.
If the $i$-th neighbour of $v_X$ is in $R_X$ then
let $X_{i}(c,c')$ be the edge-boundary
pair consisting of

\begin{itemize}
\item The region $R_{X_{i}(c,c')}=R_{X}\backslash\{v_{X}\}$,
\item the distinguished boundary edge $e_{X_{i}(c,c')}=e_{i}$, and
\item the pair $(B_{X_{i}(c,c')},B_{X_{i}(c,c')}')$ of $q_{0}$-colourings
of $\mathcal{E}R_{X_{i}(c,c')}$ such that both colourings are identical
to $B_{X}$ on all edges in $\mathcal{E}R_{X_{i}(c,c')}\backslash
\{e_1,e_2,e_3\}$.
The colours of the boundary edges in $\{e_1,e_2,e_3\}$ are assigned as follows.

\begin{itemize}
\item $B_{X_{i}(c,c')}(e_{i})=c'$ and $B_{X_{i}(c,c')}'(e_{i})=c$.
\item For the boundary edge $e_j \in \{e_1,e_2,e_3\}$ such that $j < i$,
both $B_{X_{i}(c,c')}(e_{j})$ and $B_{X_{i}(c,c')}'(e_{j})$ are $c'$.
\item For the boundary edge $e_j \in \{e_1,e_2,e_3\}$ such that $j > i$,
both $B_{X_{i}(c,c')}(e_{j})$ and $B_{X_{i}(c,c')}'(e_{j})$ are $c$.
\end{itemize}
\end{itemize}
If the $i$-th neighbour of $v_X$ is in $R_X$,
recursively construct the tree $T_{X_{i}(c,c')}$ and join it to $T_{X}$ by adding
an edge with label $(1,\cdot)$ from $r_{c,c'}$ to the root of $T_{X_{i}(c,c')}$.
Note that if $v_X$ has no neighbours in $R_X$ then $r_{c,c'}$ is a leaf.
That completes the construction of $T_{X}$.

We say that an edge $e$ of $T_{X}$ is \emph{degenerate} if the second
component of its label is {}``$\cdot$''. For edges $e$ and $e'$
of $T_{X}$, we write $e\rightarrow e'$ to denote the fact that $e$
is and ancestor of $e'$. That is, either $e=e'$, or $e$ is a proper
ancestor of $e'$. Define the \emph{level} of an edge $e$ of $T_{X}$
to be the number of non-degenerate edges on the path from the root
down to, and including, $e$. Suppose that $e$ is an edge of $T_{X}$
with label $(p,v)$. We say that the \emph{weight} $w(e)$ of edge
$e$ is $p$. Also the \emph{name} $n(e)$ of edge $e$ is $v$. The
\emph{likelihood} $l(e)$ of $e$ is $\prod_{e':e'\rightarrow e}w(e)$.
The \emph{cost} $\gamma(v,T_{X})$ of a vertex $v\in R_{X}$ is $\sum_{e:n(e)=v}l(e)$.
If the region $R_{X}$ is not connected and vertex $v_{X}$ and a
vertex $v\in R_{X}$ belong to different connected components, then
there will be no edge with name $v$ in $T_{X}$ and we define $\gamma(v,T_{X})=0$.
We have the following lemma, which
is proved in~\cite{gmp-ssmfc-04} as Lemma~12.

\begin{lemma}
[{\cite[Lemma~12]{gmp-ssmfc-04}}]\label{lem:cost-in-tx}For every
edge-boundary pair $X$ there exists a coupling $\Psi_{X}$ of $\pi_{B_{X}}$
and $\pi_{B_{X}'}$ such that $\mathbb{E}[1_{\Psi_{X},v}]\leq\gamma(v,T_{X})$
for all $v\in R_{X}$. 
\end{lemma}
A key ingredient from the construction of $T_{X}$ that affects $\gamma(v,T_{X})$
is the quantity $\mathbb{E}[1_{\Psi_{X}^{\textup{min}},v_{X}}]$,
which we denote $\nu(X)$. Thus,\[
\nu(X)=\mathbb{E}[1_{\Psi_{X}^{\textup{min}},v_{X}}]=\sum_{\substack{c,c'\in Q,\\
c\neq c'}
}p_{X}^{\textup{min}}(c,c').\]

For an edge-boundary pair $X$ and an integer $d\geq1$, let $E_{d}(X)$
denote the set of level-$d$ edges in $T_{X}$, and define
$\Gamma^{d}(X)=\sum_{e\in E_{d}(X)}l(e)$.
We define $\Gamma^d(\emptyset) = 0$ for $d \geq 1$.
Equivalently, we can define $\Gamma^{d}(X)$ recursively:\begin{equation}
\Gamma^{1}(X)=\nu(X)=\sum_{\substack{c,c'\in Q,\\
c\neq c'}
}p_{X}^{\textup{min}}(c,c'),\label{eq:gamma-base-case-definition}\end{equation}
and for $d>1$ we have
\begin{equation}
\Gamma^{d}(X)=\sum_{\substack{c,c'\in Q,\\
c\neq c'}
}p_{X}^{\textup{min}}(c,c')\sum_{i=1}^{3}\Gamma^{d-1}(X_{i}(c,c')).
\label{eq:gamma-recursion-definition}
\end{equation}

\begin{lemma}
\label{lem:gamma-to-edges}
Suppose $X$ is an edge-boundary pair and $R' \subseteq R_X$.
Then there is a coupling
$\Psi_{X}$ of $\pi_{B_{X}}$ and $\pi_{B_{X}'}$
such that
$$
\sum_{v \in R'} \mathbb{E}[1_{\Psi_{X},v}]
\leq
\sum_{d \geq d(e_X, R')} \Gamma^d(X).
$$
\end{lemma}

\begin{proof}
By Lemma~\ref{lem:cost-in-tx} there is a coupling $\Psi_{X}$ of
$\pi_{B_{X}}$ and $\pi_{B_{X}'}$ such that
$\mathbb{E}[1_{\Psi_{X},v}] \leq \gamma(v,T_{X})$
for $v \in R_{X}$. Thus,
\begin{align*}
\sum_{v \in R'} \mathbb{E}[1_{\Psi_{X},v}]
&\leq
\sum_{v \in R'} \gamma(v,T_{X})
\leq
\sum_{v \in R'} \sum_{e:n(e)=v} l(e) \\
&\leq
\sum_{d \geq d(e_{X},R')} \sum_{e \in E_{d}(X)} l(e)
\leq
\sum_{d \geq d(e_X, R')} \Gamma^d(X).
\end{align*}
\end{proof}

\section{Exponential decay of $\Gamma^{d}(X)$}

\label{sec:gamma-exp-decay}

Suppose $X$ is an edge-boundary pair.
Let $B$ be the colouring of $\mathcal{E}R_X$
such that $B(e) = B_X(e)$ for $e \in \mathcal{E}R_X \setminus \{e_X\}$
and $B(e_X) = 0$.
For $i \in Q$, we define $n_i(X)$ to be the number of proper $q$-colourings $\sigma$
in $\Omega_{R_X}(B)$ such that $\sigma(v_X) = i$.
For $i, i' \in Q$, we define
$N_{i,i'}(X) = \sum_{j \in Q \setminus \{i,i'\}} n_j(X)$
and
$$
\mu_{i,i'}(X)=\frac{n_{i}(X)}{n_{i}(X)+N_{i,i'}(X)}.
$$
Suppose
$c = B_{X}(e_{X})$ and $c' = B_{X}'(e_{X})$.
Then $\mu_{c,c'}(X)$ is the probability that $v_X$ receives colour $c$ in
$\pi_{B'_X}$, and
$\mu_{c',c}(X)$ is the probability that $v_X$ receives colour $c'$ in
$\pi_{B_X}$.
We now define
$\mu(X) = \max[\mu_{c,c'}(X),\; \mu_{c',c}(X)]$.


\begin{lemma}
\label{lem:mu-upper-bounds-nu}For every edge-boundary pair $X$,
$\nu(X)\leq\mu(X)$.
\end{lemma}

\begin{proof}
Let $X$ be an edge-boundary pair and suppose without loss of generality
that $B_{X}(e_{X}) = c$ and $B_{X}'(e_{X}) = c'$.
Suppose first that $\mu_{c,c'}(X)\geq\mu_{c',c}(X)$.
Then $n_c(X) \geq n_{c'}(X)$.
We define a coupling $\Psi_X$ of $\pi_{B_{X}}$ and $\pi_{B_{X}'}$ as follows.
Let $(C,C')$ be a pair of colourings drawn from $\Psi_{X}$ such that
$C$ is drawn from $\pi_{B_{X}}$ and $C'$ from $\pi_{B_{X}'}$.
We have $\Pr[C(v_X)=c] = 0$, $\Pr[C'(v_X)=c'] = 0$,
$\Pr[C'(v_X)=c] \geq \Pr[C(v_X)=c']$ and
$\Pr[C'(v_X)=i] \leq \Pr[C(v_X)=i]$ for $i \in Q \setminus \{c, c'\}$.
We pair up colourings in $(C,C')$ such that
$C(v_X) = C'(v_X)$ when $C'(v_X) = i$ for $i \in Q \setminus \{c, c'\}$.
Then $C(v_X) \neq C'(v_X)$ only when $C'(v_X) = c$.
Thus, $\Pr[C(v_X) \neq C'(v_X)] = \mu_{c,c'}(X)$
and
$\nu(X) \leq \mu_{c,c'}(X)$.
Suppose second that $\mu_{c',c}(X) \geq \mu_{c,c'}(X)$.
Similarly to above,
$\nu(X) \leq \mu_{c',c}(X)$.
Thus, $\nu(X) \leq \mu(X)$.

\end{proof}

Suppose $X$ is an edge-boundary pair and $c=B_{X}(e_{X})$ and
$c'=B_{X}'(e_{X})$. In order to obtain sufficiently good upper bounds
on $\nu(X)$ we use the previous lemma together with Lemma~\ref{lem:convexity-lemma}
below, which we first describe in words.
Suppose we want to upper-bound $\mu_{c,c'}(X)$. The idea is
to pick a subregion $R'\subseteq R_{X}$ that contains vertex $v_{X}$.
Then we compute the maximum value of $\mu_{c,c'}$ for that subregion,
where we maximise over colourings of the boundary of $R'$ that are
identical to $B_{X}$ on the overlapping boundary edges
$\mathcal{E}R_{X}\cap\mathcal{E}R'$.
This maximum value is an upper bound on $\mu_{c,c'}(X)$. Note that
Goldberg, Martin and Paterson~\cite[Lemma~13]{gmp-ssmfc-04} gave
a similar lemma in terms of $\mu(X)$. However, in this paper it is
crucial to be precise about the order of the colours $c$ and $c'$
in $\mu_{c,c'}(X)$.

\begin{lemma}
\label{lem:convexity-lemma}Suppose that $X$ is an edge-boundary
pair and let $c=B_{X}(e_{X})$, $c'=B_{X}'(e_{X})$. Let $R'$ be
any subset of $R_{X}$ which includes $v_{X}$. Let $S$ be the set
of edge-boundary pairs $X'$ such that $R_{X'}=R'$, the distinguished
edge $e_{X'}=e_{X}$, and for the boundary colourings $B_{X'}$ and
$B_{X'}'$ we have $B_{X'}(e)=B_{X}(e)$ 
and $B_{X'}'(e)=B_{X}'(e)$ on $e\in\mathcal{E}R_{X}\cap\mathcal{E}R'$.
Then $\mu_{c,c'}(X)\leq\max_{X'\in S}\mu_{c,c'}(X')$.
\end{lemma}
\begin{proof}
Let $X$ be an edge-boundary pair and let $c=B_{X}(e_{X})$ and $c'=B'_{X}(e_{X})$.
For a subregion $R'\subseteq R_{X}$ that contains $v_{X}$, let $H=R_{X}\backslash R'$.
For $i\in Q\backslash\{c\}$ and $\theta\in\Omega_{H}$, let $n_{i,\theta}$
denote the number of colourings in $\Omega_{R_{X}}(B_{X})$ which
colour $v_{X}$ with colour $i$ and $H$ with colouring $\theta$.
For $\theta\in\Omega_{H}$, let $n_{c,\theta}$ denote the number
of colourings in $\Omega_{R_{X}}(B_{X}')$ which colour $v_{X}$ with
colour $c$ and $H$ with colouring $\theta$.
Let $N_{c,c',\theta}=\sum_{i\in Q\backslash(c,c')}n_{i,\theta}$.
Then\begin{eqnarray*}
\mu_{c,c'}(X) & = & \frac{n_{c}(X)}{n_{c}(X)+N_{c,c'}(X)}\,\,=
\,\,\frac{\sum_{\theta\in\Omega_{H}}n_{c,\theta}}{\sum_{\theta\in\Omega_{H}}
(n_{c,\theta}+N_{c,c',\theta})}\\
 & \leq & \max_{\theta\in\Omega_{H}}\frac{n_{c,\theta}}{n_{c,\theta}+N_{c,c',\theta}}
\,\,\leq\,\,\max_{X'\in S}\mu_{c,c'}(X').\end{eqnarray*}
To see the last inequality, take any $\theta\in\Omega_{H}$ and construct
the edge-boundary pair $X'$ in $S$ with the following parameters:
$R_{X'}=R'$, $B_{X'}=B_{X}$ on $\mathcal{E}R_{X}\cap\mathcal{E}R'$
and $B_{X'}'=B_{X}'$ on $\mathcal{E}R_{X}\cap\mathcal{E}R'$. For
each boundary edge $e\in\mathcal{E}R'$ such that $e\notin\mathcal{E}R_{X}$,
let $B_{X'}(e)=B_{X'}'(e)=\theta(v)$, where vertex $v\in H$ is the endpoint of
$e$ in $\partial R'$. Now,\[
\frac{n_{c,\theta}}{n_{c,\theta}+N_{c,c',\theta}}=
\frac{n_{c}(X')}{n_{c}(X')+N_{c,c'}(X')}=\mu_{c,c'}(X').\]

\end{proof}

\subsection{Extended regions}

It will be convenient to introduce the notion of an \emph{extended
region} $\exR$, which is a region with the following additional information:
(i) Every vertex in $\exR$ is labelled either {}``in'' or {}``out'', and
(ii) one of the boundary edges of $\exR$ is referred to as the designated
edge. 

\begin{figure}[tp]
\centering
(a)~~\includegraphics[scale=\myscaling]{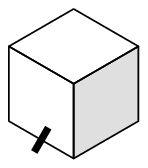}
\hspace{14mm}(b)~~\includegraphics[scale=\myscaling]{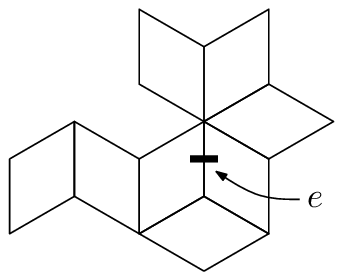}
\hspace{14mm}(c)~~\includegraphics[scale=\myscaling]{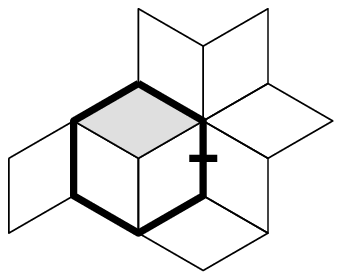}

\smallskip

\caption{\label{fig:extended-region-example}(a) An extended region $\exR$.
A non-shaded vertex is labelled ``in'', and a shaded vertex is labelled ``out''.
(b)~A region $R$ with a marked edge $e \in E(R)$. (c)~We see that $\exR$
matches $R$ with respect to edge $e$ in $R$.}

\end{figure}

An extended region $\exR$ and a region $R$ are \emph{matching}
with respect to an edge $e\in E(R)$ if there is a way of overlapping
$R$ with $\exR$ such that the designated edge of $\exR$ coincides
with the edge $e$, and every vertex that is labelled ``in'' in
$\exR$ coincides with a vertex that is in $R$, and every vertex
that is labelled ``out'' in $\exR$ coincides with a vertex that
is not in $R$. When illustrating extended regions in the figures,
we let non-shaded faces represent vertices that are labelled {}``in'',
and we let shaded faces represent vertices that are labelled {}``out''.
We mark the designated boundary edge with a short and thick line segment.
Figure~\ref{fig:extended-region-example}
 illustrates how an extended region $\exR$ matches a region $R$
with respect to an edge $e$. Note that the overlapping takes place
under any rotation or reflection of the regions.

Suppose $\exR$ is an extended region.
An extended region $\exR'$ is an \emph{extended subregion} of $\exR$
if $\exR'$ is obtained from $\exR$
by removing vertices, except for the vertex that is incident to the designated edge.
The labelling of the vertices in $\exR'$
is identical to the labelling of the same vertices
in $\exR$.

\subsection{A collection $\mathcal{F}$ of edge-boundary pairs}

\begin{figure}[tp]
\centering
(a)~~~\includegraphics[scale=\myscaling]{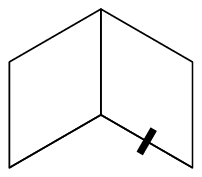}
\qquad (b)~~~\includegraphics[scale=\myscaling]{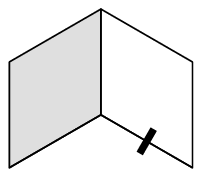}
\qquad (c)~~\includegraphics[scale=\myscaling]{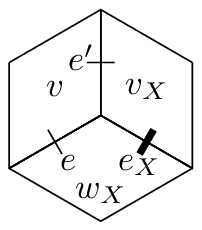}

\smallskip

\caption{\label{fig:kagome-four-sets}
(a) The extended region $\exR_{M_{(1,2)}}$.
(b) The extended region $\exR_{M_{(3,4)}}$.
(c) Labelling of vertices and edges.
}

\end{figure}

Let $\exR_{M_{(1,2)}}$ be the extended region in
Figure~\ref{fig:kagome-four-sets}(a) and
let $\exR_{M_{(3,4)}}$ be the extended region in
Figure~\ref{fig:kagome-four-sets}(b).
Let $M_{(1,2)}$ be the set of edge-boundary pairs $X$ such that $R_X$ and
$\exR_{M_{(1,2)}}$ are matching with respect to $e_X$.
Let $M_{4}$ be the set of edge-boundary pairs $X$ such that $R_X$ and
$\exR_{M_{(3,4)}}$ are matching with respect to $e_X$.
Let $X$ be an edge-boundary pair and
suppose
$c = B_X(e_X)$ and $c' = B'_X(e_X)$.
Let $v$ be the vertex that is a neighbour to both $v_{X}$ and $w_{X}$,
let $e$  be the edge between $w_X$ and $v$, and
let $e'$ be the edge between $v$ and $v_X$
(see Figure~\ref{fig:kagome-four-sets}(c)).
The three sets $M_1 \subseteq M_{(1,2)}$, $M_2 \subseteq M_{(1,2)}$
and $M_3 \subseteq M_4$ of edge-boundary pairs are defined as follows.
\begin{itemize}
\item $X \in M_1$ if $v \in R_X$ and either
$\mu_{c,c'} \geq \mu_{c',c}$ and $B_X(e) = c$, or
$\mu_{c',c} \geq \mu_{c,c'}$ and $B_X(e) = c'$.
\item $M_2 = M_{(1,2)} \setminus M_1$.
\item $X \in M_3$ if $v \notin R_X$ and either $B_X(e') = c$ or $B_X(e') = c'$.
\end{itemize}

\begin{figure}[tp]
\centering
(a)\includegraphics[scale=\myscaling]{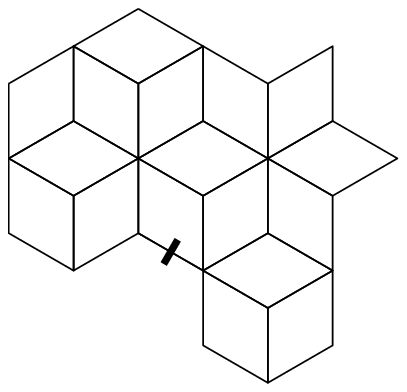}
\qquad(b)\includegraphics[scale=\myscaling]{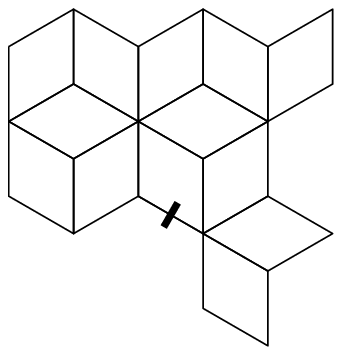}
\qquad(c)\includegraphics[scale=\myscaling]{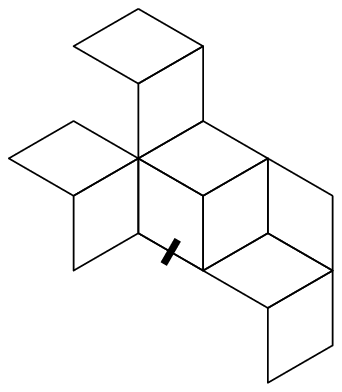}

\smallskip

\caption{\label{fig:kagome-region-f}(a) The extended region $\exR_{F}$. (a)--(c)
Three of the extended regions $\exR_{F_{1}},\dots,\exR_{F_{4720}}$.
All vertices are labelled ``in''.}
\end{figure}

Let $\exR_{F}$ be the extended region in Figure~\ref{fig:kagome-region-f}(a).
For $f\in\{1,\dots,4720\}$ we define the extended region $\exR_{F_{f}}$
such that it is an extended subregion of $\exR_{F}$.
Note that all vertices in $\exR_{F_{f}}$ are labelled ``in''.
The remark on page~\pageref{rem:why-a-f} explains why we define exactly
these 4720 extended
regions.
Due to the
large number of extended regions we only illustrate three of them
here (Figure~\ref{fig:kagome-region-f}). For $f\in\{1,\dots,4720\}$,
let $F_{f}$ be the set of edge-boundary pairs $X$ such that $R_{X}$
and $\exR_{F_{f}}$ are matching with respect to edge $e_{X}$.
For $m \in \{1,\dots,4\}$,
let $F_{f,m}=F_{f}\cap M_{m}$.
Let
$\mathcal{F}$ be the collection of all sets $F_{f,m}$.
One of the extended regions $\exR_{F_{1}},\dots,\exR_{F_{4720}}$
is defined to contain only the single vertex that is incident to the
designated edge. Hence any edge-boundary pair $X$ is guaranteed to
belong to at least one of the sets in $\mathcal{F}$.
Note that many of the sets $F_{f,m}$ are empty.
For instance, if $\exR_{F_i}$ is the extended region in
Figure~\ref{fig:kagome-region-f}(b) for some $i \in \{1,\dots,4720\}$ then obviously no
edge-boundary pair $X$ can belong to both $F_i$ and $M_4$.
Hence $F_{i,4} = \emptyset$.

\subsection{The constants $\mu_{F_{f,m}}$}
\label{sec:the-constants}

For $f \in \{1, \dots, 4720\}$ we define $F'_f$ to be the set of edge-boundary pairs $X$
such that the vertices of $R_X$ are exactly those of $\exR_{F_f}$,
$e_X$ is the designated edge of $\exR_{F_f}$,
$B_X(e_X) = 1$, $B'_X(e_X) = 2$, and the number of colours $q = 5$.
For an edge-boundary pair $X \in F'_f$,
let $v$ be the vertex that is a neighbour to both $v_{X}$ and $w_{X}$,
let $e$  be the edge between $w_X$ and $v$, and
let $e'$ be the edge between $v$ and $v_X$
(see Figure~\ref{fig:kagome-four-sets}(c)).
Suppose first that $\exR_{M_{(1,2)}}$ is an extended subregion of $\exR_{F_f}$.
Then
we define $F'_{f,1} \subseteq F'_f$ to be the set of edge-boundary pairs $X \in F'_f$
such that $B_X(e) = 1$,
we define $F'_{f,2} \subseteq F'_f$ to be the set of edge-boundary pairs $X \in F'_f$
such that $B_X(e) = 2$,
and we define $F'_{f,3} = F'_{f,4} = \emptyset$.
Suppose second that $\exR_{M_{(1,2)}}$ is not an extended subregion of $\exR_{F_f}$.
Then
we define $F'_{f,1} = F'_{f,2} = \emptyset$,
we define $F'_{f,3} \subseteq F'_f$ to be the set of edge-boundary pairs $X \in F'_f$
such that either $B_X(e') = 1$ or $B_X(e') = 2$, and we define
$F'_{f,4} = F'_f$.
Now, for $f \in \{1, \dots, 4720\}$ and $m \in \{1, \dots, 4\}$, we define
$$
\mu_{F_{f,m}} = \max_{X \in F'_{f,m}} \mu_{1,2}(X)
$$
if $F'_{f,m} \neq \emptyset$,
and $\mu_{F_{f,m}} = 0$ if $F'_{f,m} = \emptyset$.

\begin{lemma}
\label{lem:kagome-mu-values}
Suppose $q=5$, $f \in \{1, \dots, 4720\}$, $m \in \{1,\dots,4\}$
and $F_{f,m} \neq \emptyset$.
Then
$\nu(X) \leq \mu_{F_{f,m}}$
for every edge-boundary pair $X \in F_{f,m}$.
\end{lemma}

\begin{proof}
Suppose $f \in \{1, \dots, 4720\}$ and $m \in \{1,\dots,4\}$
such that
$F_{f,m} \neq \emptyset$.
Let $X$ be an edge-boundary pair in $F_{f,m}$.
Let $v$ be the vertex that is a neighbour to both $v_{X}$ and $w_{X}$,
let $e$  be the edge between $w_X$ and $v$, and
let $e'$ be the edge between $v$ and $v_X$
(see Figure~\ref{fig:kagome-four-sets}(c)).
From Lemma~\ref{lem:mu-upper-bounds-nu} we have that $\nu(X) \leq \mu(X)$.
In order to upper-bound $\mu(X)$ we may assume
without loss of generality that
$B_X(e_X) = 1$ and $B'_X(e_X) = 2$.

Suppose first that $m = 1$.
Without loss of generality we may assume that $B_X(e) = 1$ and
hence $\mu_{1,2}(X) \geq \mu_{2,1}(X)$.
Then $\mu(X) = \mu_{1,2}(X)$.
Let $R'$ be the subset of $R_X$ such that
the vertices of $R'$ are exactly those of $\exR_{F_f}$.
Let $S$ be the set
of edge-boundary pairs $X'$ such that $R_{X'}=R'$, the distinguished
edge $e_{X'}=e_{X}$, and for the boundary colourings $B_{X'}$ and
$B_{X'}'$ we have
$B_{X'}(e'') = B_{X}(e'')$
and $B_{X'}'(e'')=B_{X}'(e'')$ on $e'' \in \mathcal{E}R_{X}\cap\mathcal{E}R'$.
Note that $S \subseteq F'_{f,m}$. We have
$$
\mu_{1,2}(X) \leq \max_{X'\in S}\mu_{1,2}(X')
\leq \max_{X'\in F'_{f,m}}\mu_{1,2}(X')
= \mu_{F_{f,m}},
$$
where the first inequality is from Lemma~\ref{lem:convexity-lemma}.

Suppose second that $m = 2$.
Without loss of generality we may assume that
$B_X(e) = 2$ and
hence $\mu_{1,2}(X) \geq \mu_{2,1}(X)$.
Proceeding as above we see that $\mu(X) \leq \mu_{F_{f,m}}$.
Now suppose $m = 3$.
Without loss of generality we may assume that
$B_X(e') = 1$ or $B_X(e') = 2$ and
$\mu_{1,2}(X) \geq \mu_{2,1}(X)$.
Proceeding as above we see that $\mu(X) \leq \mu_{F_{f,m}}$.
Lastly, for $m = 4$ we make no assumption on the colour of edge $e'$
and again we see that $\mu(X) \leq \mu_{F_{f,m}}$.

\end{proof}

\subsection{A collection $\mathcal{A}$ of edge-boundary pairs}

\begin{figure}[tp]
\centering
(a)\includegraphics[scale=\myscaling]{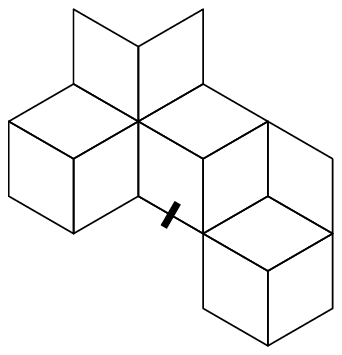}
\qquad\qquad(b)\includegraphics[scale=\myscaling]{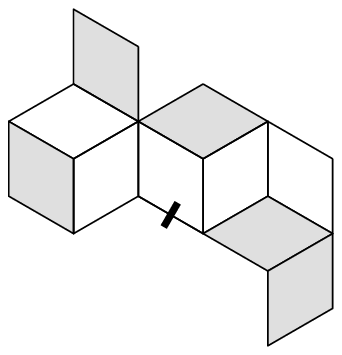}
\qquad\qquad(c)\includegraphics[scale=\myscaling]{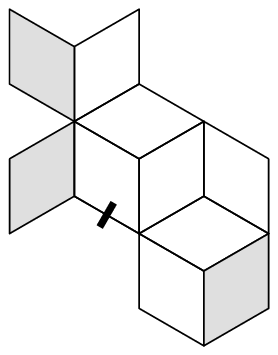}

\smallskip

\caption{\label{fig:kagome-region-a}(a) The extended region $\exR_{A}$. (a)--(c)
Three of the extended regions $\exR_{A_{1}},\dots,\exR_{A_{342}}$.}

\end{figure}

Let $\exR_{A}$ be the extended region in Figure~\ref{fig:kagome-region-a}(a).
For $a \in \{1,\dots,342\}$ we define the extended region $\exR_{A_{a}}$
to be a subregion of $\exR_{A}$.%
 The extended regions are defined such that for any edge-boundary
pair $X$, the region $R_{X}$ matches exactly one of $\exR_{A_{1}},\dots,\exR_{A_{342}}$
with respect to edge $e_{X}$.
The remark on page~\pageref{rem:why-a-f} explains why we define exactly
these 342 extended
regions.
In Figure~\ref{fig:kagome-region-a}
we illustrate three of the 342 extended regions. For $a \in \{1,\dots,342\}$,
let $A_{a}$ be the set of edge-boundary pairs $X$ such that $R_{X}$
matches $\exR_{A_{a}}$ with respect to edge $e_{X}$. Furthermore,
for $m \in \{1,\dots,4\}$ we define $A_{a,m}=A_{a}\cap M_{m}$, and define
$\mathcal{A}$ to be the collection of all sets $A_{a,m}$.
Note that many of the sets $A_{a,m}$ are empty.

\subsection{Exponential decay}

A set
$
\calS \subseteq \calA \times \calF \times \calA \times \calA \times \calA
$
is called an \emph{$(\calA, \calF)$-set}
if the following is true about $\calS$:
For
every set $A_{a,m} \in \calA$,
every edge-boundary pair
$X \in A_{a,m}$, and
every two distinct colours
$c, c' \in Q$ such that $p_{X}^{\textup{min}}(c,c') > 0$,
there is a 5-tuple
$(A_{a,m}, F_{f,m},\nolinebreak[3] A_{a_1,m_1},\nolinebreak[3] A_{a_2,m_2}, A_{a_3,m_3})$
in $\calS$, such that
$X \in F_{f,m}$, and
for $i \in \{1,2,3\}$
the edge-boundary pair $X_{i}(c,c')$ constructed recursively in the tree $T_{X}$
belongs to $A_{a_i,m_i}$.
For values of $i$ such that $X_{i}(c,c') = \emptyset$, $A_{a_i,m_i} = \emptyset$.

Suppose 
$\epsilon \in (0,1)$ is a constant.
An $(\calA, \calF)$-set $\calS$
is \emph{good with respect to $\epsilon$}
if the following is true:
For $i \in \{1, \dots, 342\}$ and $j \in \{1,\dots,4\}$
there is a constant
$\alpha_{A_{i,j}}$ such that
$\alpha_{A_{i,j}} \geq 1/(1-\epsilon)$ if $A_{i,j} \neq \emptyset$ and
$\alpha_{A_{i,j}} \geq 0$ if $A_{i,j} = \emptyset$,
and
for every 5-tuple
$(A_{a,m}, F_{f,m},\nolinebreak[3] A_{a_1,m_1},\nolinebreak[3] A_{a_2,m_2}, A_{a_3,m_3})$
in $\calS$,
\begin{equation}
\label{eq:good}
\mu_{F_{f,m}}
(\alpha_{A_{a_1,m_1}} +
\alpha_{A_{a_2,m_2}} +
\alpha_{A_{a_3,m_3}})
\leq \alpha_{A_{a,m}}(1-\epsilon).
\end{equation}

\begin{lemma}
\label{lem:gamma-exp-decay}
Suppose $q = 5$, $\epsilon \in (0,1)$ is a constant,
and $\calS$ is an $(\calA,\calF)$-set that is good with respect to $\epsilon$.
Then there is a constant
$\alpha \geq 0$ such that $\Gamma^{d}(X) \leq \alpha(1-\epsilon)^{d}$
for all edge-boundary pairs $X$.
\end{lemma}

\begin{proof}
Since $\calS$ is good with respect to $\epsilon$, there are constants
$\alpha_{A_{a,m}}$,
$a \in \{1,\dots,342\}$ and $m \in \{1,\dots,4\}$,
such that
Equation~(\ref{eq:good}) is satisfied for every 5-tuple in $\calS$.
For $A_{a,m} \in \calA$,
let $\Gamma^{d}(A_{a,m})$ denote
the maximum of $\Gamma^{d}(X)$ over all $X\in A_{a,m}$.
Remember $\Gamma^{d}(\emptyset)=0$
for $d \geq 1$.
In order to show that
there is a constant $\alpha$ such that
$\Gamma^{d}(X) \leq \alpha(1-\epsilon)^{d}$
for every edge-boundary pair $X$,
we will show that $\Gamma^{d}(A_{a,m}) \leq \alpha_{A_{a,m}}(1-\epsilon)^{d}$
for every non-empty set $A_{a,m} \in \calA$.
Then we let $\alpha$ be the maximum of
$\alpha_{A_{a,m}}$ over all
$a \in \{1,\dots,342\}$ and $m \in \{1,\dots,4\}$.
Note that any edge-boundary pair $X$ belongs to at
least one of the sets in $\calA$.

Consider any non-empty set $A_{a,m} \in \calA$
and any edge-boundary pair $X \in A_{a,m}$.
We are going to show that $\Gamma^d(X) \leq \alpha_{A_{a,m}}(1-\epsilon)$
by induction on $d$.
We start with the base case $d=1$.
Since $\alpha_{A_{a,m}} \geq 1/(1-\epsilon)$,
we have
$$
\Gamma^{1}(X) = \nu(X) \leq \mu(X) \leq 1 \leq \alpha_{A_{a,m}} (1-\epsilon),
$$
where the first inequality is from Lemma~\ref{lem:mu-upper-bounds-nu}.
Now consider the inductive step.
We repeat Equation~(\ref{eq:gamma-recursion-definition}):
\begin{equation}
\label{eq:exp-decay-gamma-def}
\Gamma^{d}(X)=\sum_{\substack{c,c'\in Q,\\
c \neq c'}}
p_{X}^{\textup{min}}(c,c')
\sum_{i=1}^{3}
\Gamma^{d-1}(X_{i}(c,c')),
\end{equation}
where $X_{i}(c,c')$ is the edge-boundary pair constructed recursively
in the tree $T_{X}$.
Here $Q = \{1,\dots,5\}$.
For every two distinct colours $c, c' \in Q$
such that
$p_{X}^{\textup{min}}(c,c') > 0$,
we know that there is a 5-tuple
$(A_{a,m}, F_{f,m},\nolinebreak[3] A_{a_1,m_1},\nolinebreak[3] A_{a_2,m_2}, A_{a_3,m_3})$
in $\calS$
such that
$X\in F_{f,m}$ and
$X_{i}(c,c') \in A_{a_{i},m_{i}}$, where $i \in \{1,2,3\}$.
If the $i$-th neighbour of $v_{X}$ is not in
$R_{X}$ then we have
$A_{a_i,m_i} = \emptyset$.
By the induction hypothesis we have
\begin{equation}
\label{eq:exp-decay-sum-gamma}
\sum_{i=1}^{3}\Gamma^{d-1}(X_{i}(c,c'))\leq
\sum_{i=1}^{3}\Gamma^{d-1}(A_{a_{i},m_{i}})\leq
\sum_{i=1}^{3}\alpha_{A_{a_{i},m_{i}}}(1-\epsilon)^{d-1}.
\end{equation}
Using Equation~(\ref{eq:exp-decay-gamma-def})
with Equation~(\ref{eq:exp-decay-sum-gamma}) gives
\begin{align*}
\Gamma^{d}(X)\, & \leq\sum_{\substack{c,c'\in Q,\\
c\neq c'}
}p_{X}^{\textup{min}}(c,c')\sum_{i=1}^{3}\alpha_{A_{a_{i},m_{i}}}(1-\epsilon)^{d-1} =
\nu(X)\sum_{i=1}^{3}\alpha_{A_{a_{i},m_{i}}}(1-\epsilon)^{d-1}\\
&
\leq \mu_{F_{f,m}} \sum_{i=1}^{3}\alpha_{A_{a_{i},m_{i}}}(1-\epsilon)^{d-1}
\leq \alpha_{A_{a,m}}(1-\epsilon)^{d},
\end{align*}
where
$\nu(X) \leq \mu_{F_{f,m}}$ is from Lemma~\ref{lem:kagome-mu-values},
and the last inequality follows from Equation~(\ref{eq:good}).
\end{proof}

The next lemma is proved by computer assistance and we will explain the details in
Section~\ref{sec:computational-part}.

\begin{lemma}
\label{lem:gamma-exp-decay-final}
Suppose $q = $ and $\epsilon = 1/1000$.
Then $\Gamma^d(X) \leq 5(1-\epsilon)^d$
for every edge-boundary pair $X$.
\end{lemma}

\begin{proof}
In order to prove this lemma we use computer assistance.
The computerised steps are to first calculate all the constants $\mu_{F_{f,m}}$,
then generate an $(\calA,\calF)$-set $\calS$ that is good with respect to
$\epsilon = 1/1000$.
This last step is broken into the following steps.
First we generate an $(\calA,\calF)$-set $\calS$.
Then, for every 5-tuple
$(A_{a,m}, F_{f,m},\nolinebreak[3] A_{a_1,m_1},\nolinebreak[3] A_{a_2,m_2}, A_{a_3,m_3})$
in $\calS$, we add the inequality
$$
\mu_{F_{f,m}}
(\alpha_{A_{a_1,m_1}} +
\alpha_{A_{a_2,m_2}} +
\alpha_{A_{a_3,m_3}})
\leq \alpha_{A_{a,m}}(1-\epsilon)
$$
to a linear program. The unknowns in this linear program are the variables
$\alpha_{A_{a,m}}$.
A solution to the linear program is found with $\alpha_{A_{a,m}} \in [2,5]$
for $A_{a,m} \neq \emptyset$ and 
$\alpha_{A_{a,m}} \in [0,5]$
for $A_{a,m} = \emptyset$.
Hence $\calS$ is good with respect to $\epsilon$.
By Lemma~\ref{lem:gamma-exp-decay} it follows that
$\Gamma^d(X) \leq \alpha(1-\epsilon)^d$ for every edge-boundary pair $X$,
where $\alpha \geq 0$ is a constant.
From the proof of Lemma~\ref{lem:gamma-exp-decay} we see that we can choose
$\alpha$ to be the maximum of all $\alpha_{A_{a,m}}$, which is~5.
\end{proof}


\begin{remark*}
\label{rem:why-a-f}
One probably asks why the sets in $\calA$ and $\calF$ are the sets we use to prove mixing.
The sets in $\calA$ and $\calF$, or the extended regions $\exR_{A_i}$ and $\exR_{F_i}$
to be more precise, have arisen from a lengthy process of trial and error and experiments.
One part of the proof of Lemma~\ref{lem:gamma-exp-decay-final} above
is to find a solution to a linear program.
If the values
$\mu_{F_{f,m}}$ are too large then there will be no solution to this linear program.
In order to obtain smaller values $\mu_{F_{f,m}}$ we must increase the size of the
regions $\exR_{F_i}$.
Small extended regions $\exR_{A_i}$
contain only little information about which vertices
are in and not in the region $R_X$ for an edge-boundary pair $X \in A_i$.
In particular, with small regions $\exR_{A_i}$ we quickly lose information
about which vertices
are in and not in the regions $R_{X_i(c,c')}$ for the recursively
constructed edge-boundary pairs $X_i(c,c')$.
Thus, too small extended regions $\exR_{A_i}$ will result in a
linear program that is too small and has no solution.
We started with a few small extended regions $\exR_{A_i}$ and $\exR_{F_i}$ and
slowly increased the sizes of them until we obtained a
linear program that could be successfully solved.
We let the regions grow in a way that seemed reasonable based on
experiments and intuition.
\end{remark*}

\section{Strong spatial mixing}
\label{sec:exp-decay-vertices-and-ssm}


\begin{lemma}
\label{lem:edges-to-vertices}
Suppose $\epsilon = 1/1000$ and $q = 5$.
Suppose $\vbp$ is a vertex-boundary pair and $R' \subseteq R_\vbp$.
Then there is a coupling $\Psi_{\vbp}$
of $\pi_{\mathcal{B}_{X}}$ and $\pi_{\mathcal{B}'_{X}}$ such that
$$
\sum_{v \in R'} \mathbb{E}[1_{\Psi_{\vbp}, v}]
\leq
\frac{30}{\epsilon (1-\epsilon)}(1-\epsilon)^{d(w_{\vbp}, R')}.
$$
\end{lemma}

\begin{proof}
First suppose that $w_{\vbp}$ has a neighbour $y\notin R_{\vbp}$.
Let $k=|E|\leq3$,
where $E=\{e_{1},\dots,e_{k}\}\subseteq\mathcal{E}R_{\vbp}$ is the
set of boundary edges incident to $w_{\vbp}$. Label the edges in
$E$ clockwise around $w_{\vbp}$ so that edge $(w_{\vbp},y)$ appears
between edge $e_{k}$ and $e_{1}$ when traversing edges around $w_{\vbp}$
in clockwise direction. This guarantees that $e_{i}$ and $e_{j}$
are adjacent only if $i$ and $j$ differ by~1.

For $i=1,\dots,k$, let $X_{i}$ be the edge-boundary pair consisting
of region $R_{X_{i}}=R_{\vbp}$, the distinguished edge $e_{X_{i}}=e_{i}$,
and boundary colourings $B_{X_{i}}$ and $B_{X_{i}}'$. For every
boundary edge $e=(w,v)\in\mathcal{E}R_{X_{i}}\backslash E$, where
$w\in\partial R_{\vbp}$, we have $B_{X_{i}}(e)=B_{X_{i}}'(e)=\mathcal{B}_{\vbp}(w)$.
The colours of the edges in $E$ are assigned as follows.
\begin{itemize}
\item $B_{X_{i}}(e_{j})=B_{X_{i}}'(e_{j})=\mathcal{B}_{\vbp}'(w_{\vbp})$
for $j=1,\dots,i-1$,
\item $B_{X_{i}}(e_{j})=\mathcal{B}_{\vbp}(w_{\vbp})$ and
$B_{X_{i}}'(e_{j})=\mathcal{B}_{\vbp}'(w_{\vbp})$
for $j=i$, and
\item $B_{X_{i}}(e_{j})=B_{X_{i}}'(e_{j})=\mathcal{B}_{\vbp}(w_{\vbp})$
for $j=i+1,\dots,k$.
\end{itemize}
By Lemma~\ref{lem:gamma-to-edges} there is a coupling
$\Psi_{i}$ of $\pi_{B_{X_{i}}}$ and $\pi_{B_{X_{i}}'}$
such that
\begin{equation}
\label{eq:psi-i-bound}
\sum_{v \in R'} \mathbb{E}[1_{\Psi_i,v}]
\leq
\sum_{d \geq d(e_{X_i}, R')} \Gamma^d(X_i).
\end{equation}

Let $\Psi_{\vbp}$ be the coupling of $\pi_{\mathcal{B}_{\vbp}}$
and $\pi_{\mathcal{B}_{\vbp}'}$ defined by composing the couplings
$\Psi_{1},\dots,\Psi_{k}$. More precisely, in order to choose a pair
$(\sigma_{0},\sigma_{k})$ of colourings from $\Psi_{\vbp}$, first
draw the pair $(\sigma_{0},\sigma_{1})$ from $\Psi_{1}$. Say $\sigma_{0}=x_{0}$
and $\sigma_{1}=x_{1}.$ Then choose the pair $(\sigma_{1},\sigma_{2})$
from the conditional distribution $\Psi_{2}$, conditioned on $\sigma_{1}=x_{1}$.
Say $\sigma_{2}=x_{2}$. Then choose the pair $(\sigma_{2},\sigma_{3})$
from the conditional distribution $\Psi_{3}$, conditioned on $\sigma_{2}=x_{2}$,
and so on. Hence, $\sigma_{0}$ is drawn from $\pi_{B_{X_{1}}}=\pi_{\mathcal{B}_{\vbp}}$
and $\sigma_{k}$ is drawn from $\pi_{B_{X_{k}}'}=\pi_{\mathcal{B}_{\vbp}'}$.
By the construction of the coupling $\Psi_{\vbp}$ it follows that
if the colour of a vertex $v\in R_{\vbp}$ differs in a pair $(\sigma_{0},\sigma_{k})$
drawn from $\Psi_{\vbp}$ then it must differ in at least one of the
pairs $(\sigma_{i-1},\sigma_{i})$ drawn from $\Psi_{i}$, where $i=1,\dots,k$.
Using Equation~(\ref{eq:psi-i-bound}) and Lemma~\ref{lem:gamma-exp-decay-final} we have
\begin{align*}
\sum_{v \in R'} \mathbb{E}[1_{\Psi_{\vbp},v}]
&\leq
\sum_{v \in R'}
\sum_{i=1}^{k}
\mathbb{E}[1_{\Psi_{i},v}]
=
\sum_{i=1}^{k}
\sum_{v \in R'}
\mathbb{E}[1_{\Psi_{i},v}]\\
&\leq
\sum_{i=1}^{k}
\sum_{d \geq d(e_{X_i}, R')} \Gamma^d(X_i)
\leq
\sum_{i=1}^{k}
\sum_{d \geq d(w_\vbp, R')} 5(1-\epsilon)^{d}\\
&=
\sum_{i=1}^{k}\frac{5}{\epsilon}
(1-\epsilon)^{d(w_{\vbp}, R')}
\leq
\frac{15}{\epsilon}(1-\epsilon)^{d(w_{\vbp}, R')}.
\end{align*}

Now suppose all neighbours of $w_{\vbp}$ are in $R_{\vbp}$.
Breaking the discrepancy at
vertex $w_{\vbp}$
into edge-boundary pairs $X_{i}$ as above is not possible because
the induced edge-boundary pairs are not valid with respect to the
colouring of adjacent boundary edges.

Let $u\in R_{\vbp}$ be a neighbour of $w_{\vbp}$.
Suppose $u \notin R'$.
Let
$R_{\vbp,u}=R_{\vbp}\backslash\{u\}$
be the region $R_{\vbp}$ after removing vertex $u$. For $c\in Q$,
let $\mathcal{B}_{\vbp,c}$ be the colouring of the vertex-boundary
$\partial R_{\vbp,u}$ such that for all $v\in\partial R_{\vbp}\cap\partial R_{\vbp,u}$,
$\mathcal{B}_{\vbp,c}(v)=\mathcal{B}_{\vbp}(v)$, and $\mathcal{B}_{\vbp,c}(u)=c$
(if $u\in\partial R_{\vbp,u}$). Similarly, for $c'\in Q$, let $\mathcal{B}_{\vbp,c'}'$
be the colouring of the vertex-boundary $\partial R_{\vbp,u}$ such
that for all $v\in\partial R_{\vbp}\cap\partial R_{\vbp,u}$,
$\mathcal{B}_{\vbp,c'}'(v)=\mathcal{B}_{\vbp}'(v)$,
and $\mathcal{B}_{\vbp,c'}'(u)=c'$. Note that the colourings $\mathcal{B}_{\vbp,c}$
and $\mathcal{B}_{\vbp,c'}'$ can differ on up to two vertices, namely on
vertex $w_{\vbp}$ and $u$. We break the difference in the (up to)
two vertices $w_{\vbp}$ and $u$ on the boundary $\partial R_{\vbp,u}$
into differences in the edges that bound them.

Let $k=|E|\leq6$, where $E=\{e_{1},\dots,e_{k}\}\subseteq\mathcal{E}R_{\vbp,u}$
is the set of boundary edges incident to $w_{\vbp}$ or $u$. Label
the edges in $E$ clockwise around $w_{\vbp}$ and $u$ so that $e_{k}$
and $e_{1}$ are not adjacent. Such a labelling is always possible
since $w_{\vbp}$ and $u$ are neighbours. This guarantees that $e_{i}$
and $e_{j}$ are only adjacent if $i$ and $j$ differ by~1.

Let $c\in Q$ and $c'\in Q$ be two (not necessarily different) colours.
Similarly to above, for $i=1,\dots,k$, let $X_{i}$ be the edge-boundary
pair consisting of region $R_{X_{i}}=R_{\vbp,u}$, the distinguished
edge $e_{X_{i}}=e_{i}$, and boundary colourings $B_{X_{i}}$ and
$B_{X_{i}}'$. The colourings $B_{X_{i}}$ and $B_{X_{i}}'$ are defined
similarly to above, as a sequence of colourings differing only on
the distinguished edge $e_{i}$. That is, for a boundary edge
$e=(w,v)\in\mathcal{E}R_{\vbp,u}$,
where $w\in\partial R_{\vbp,u}$, we have $B_{X_{1}}(e)=\mathcal{B}_{\vbp,c}(w)$
and $B_{X_{k}}'(e)=\mathcal{B}_{\vbp,c'}'(w)$.
Let $\Psi_{i}$ be
a coupling of $\pi_{B_{X_{i}}}$ and $\pi_{B_{X_{i}}'}$ such that
Equation~(\ref{eq:psi-i-bound}) is satisfied,
which possible due to Lemma~\ref{lem:gamma-to-edges}.
We now construct a coupling $\Psi_{\vbp}$
of $\pi_{\mathcal{B}_{\vbp}}$ and $\pi_{\mathcal{B}_{\vbp}'}$ in
the following way.

Let $\Psi_{\vbp}'$ be \emph{any} coupling of $\pi_{\mathcal{B}_{\vbp}}$
and $\pi_{\mathcal{B}_{\vbp}'}$. Let $(C,C')$ be the random variable
corresponding to the pair of colourings drawn from $\Psi_{\vbp}$
(yet to be constructed). We will choose the colour of $u$ in $C$
and $C'$ according to $\Psi_{\vbp}'$. Let $c$ and $c'$ be the
colour of $u$ drawn from $\Psi_{\vbp}'$. Let $\Psi_{\vbp,c,c'}$
be a coupling of $\pi_{\mathcal{B}_{\vbp,c}}$ and $\pi_{\mathcal{B}_{\vbp,c'}'}$.
To complete the construction of $\Psi_{\vbp}$ we colour the remaining
vertices in $R_{\vbp}$ by choosing two colourings from $\Psi_{\vbp,c,c'}$.
The coupling $\Psi_{\vbp,c,c'}$ is constructed by composing the $k$
couplings $\Psi_{X_{i}}$ as above.
We have
\begin{align*}
\sum_{v \in R'} \mathbb{E}[1_{\Psi_{\vbp},v}]
&\leq
\sum_{v \in R'}
\sum_{i=1}^{k}
\mathbb{E}[1_{\Psi_{i},v}]
=
\sum_{i=1}^{k}
\sum_{v \in R'}
\mathbb{E}[1_{\Psi_{i},v}]\\
&\leq
\sum_{i=1}^{k}
\sum_{d \geq d(e_{X_i}, R')} \Gamma^d(X_i)
\leq
\sum_{i=1}^{k}
\sum_{d \geq d(w_\vbp, R') - 1} 5(1-\epsilon)^{d}\\
&=
\sum_{i=1}^{k}\frac{5}{\epsilon}
(1-\epsilon)^{d(w_{\vbp}, R') - 1}
\leq
\frac{30}{\epsilon(1-\epsilon)}(1-\epsilon)^{d(w_{\vbp}, R')},
\end{align*}
where the $-1$ in ``$d(w_\vbp, R') - 1$'' comes from the fact that
the distance from the discrepancy edge $e_{X_i}$ to $R'$ may be one
less than $d(w_{\vbp}, R')$.
Since we sum over all distances greater than or equal to
$d(w_\vbp, R') - 1$, and $(1-\epsilon)^0 = 1$,
we note that the bound also holds when $u \in R'$.
\end{proof}




We now prove Theorem~\ref{thm:kagome-ssm} of strong spatial mixing
for $q=5$ colours.

\begin{theorem*}
[\ref{thm:kagome-ssm}, repeated]The system specified by proper 5-colourings
of the kagome lattice has strong spatial mixing.
\end{theorem*}
\begin{proof}
Consider the vertex-boundary pair $\vbp$ such that, from
Definition~\ref{def:strong-spatial-mixing}
of strong spatial mixing, we have $R_{\vbp}=R$, $\mathcal{B}_{\vbp}=\mathcal{B}$,
$\mathcal{B}_{\vbp}'=\mathcal{B}'$ and $w_{\vbp}=w$. Let $R'$ be
any subregion of $R$. The total variation distance between $\pi_{\mathcal{B},R'}$
and $\pi_{\mathcal{B}',R'}$ is upper-bounded by the probability that
$R'$ differ under any coupling $\Psi$ of $\pi_{\mathcal{B},R'}$
and $\pi_{\mathcal{B}',R'}$. This probability is upper-bounded by
$\sum_{v\in R'}\mathbb{E}[1_{\Psi},v]$. Using the coupling $\Psi_{\vbp}$
in Lemma~\ref{lem:edges-to-vertices}, we have
$$
\dtv(\pi_{\mathcal{B},R'},\pi_{\mathcal{B}',R'})=
\dtv(\pi_{\mathcal{B}_{\vbp},R'},\pi_{\mathcal{B}_{\vbp}',R'})\leq
\sum_{v\in R'}\mathbb{E}[1_{\Psi_{\vbp}},v]\leq\alpha|R'|(1-\epsilon)^{d(w_{\vbp},R')},
$$
where $\epsilon=1/1000$ and $\alpha=30/(\epsilon (1-\epsilon))$.
\end{proof}

\section{Rapid mixing}

\label{sec:rapid-mixing}The implication from strong spatial mixing
to rapidly mixing Glauber dynamics is only known to hold for graphs
of \emph{sub-exponential growth}~\cite{w-cistt-06}, meaning that
the number of vertices at distance $d$ from any vertex $v$ is sub-exponential
in $d$. This is an important property we make use of in the proof
of rapid mixing in this section. For further discussion on this topic in general,
see~\cite{gmp-ssmfc-04}, in particular \cite[Section~7.5]{gmp-ssmfc-04}.

\begin{lemma}
\label{lem:sub-exponential}Let $v\in V_{\mathcal{G}}$ be any vertex
in the kagome lattice and let $n_{d}(v)$ denote the number of vertices
at distance $d$ from $v$. Then $n_{d}(v)\in\Theta(d)$.
\end{lemma}
\begin{proof}
Recall the definition of the kagome lattice in
Section~\ref{sub:definitions-and-background},
in particular Figure~\ref{fig:kagome_lattice}. First assume that
$v\in V_{\textup{odd}}$. In order to derive lower and upper bounds
on $n_{d}(v)$, we assume without loss of generality that $v=(1,1)$
is the vertex at $x$-coordinate $1$ and $y$-coordinate $1$. Fix
any positive integer $d$.

We first derive a lower bound on $n_{d}((1,1))$. For each odd value
of $y\in\{1,\dots,d\}$, let $(x,y)$ be the vertex at distance $d$
from $(1,1)$ that is reached with the following path: $(1,1),(2,2),
\nolinebreak[1](3,3),\dots,(y,y),\nolinebreak[1](y+2,y),
\nolinebreak[1](y+4,y)\dots,(x,y)$.
Note that vertex $(y,y)\in V_{\textup{odd}}$, and from $(y,y)$ we
go as far as possible to the right. Also note that there is no path
from $(1,1)$ to $(x,y)$ that is shorter than length $d$. Thus,
there are at least $\left\lfloor d/2\right\rfloor $ vertices at distance
$d$ from $(1,1)$, and we have $n_{d}((1,1))\geq\left\lfloor d/2\right\rfloor $.

When deriving an upper bound on $n_{d}((1,1))$ we will use two claims:

\textbf{Claim 1}. For any two vertices $(x,y_{\textup{low}})$ and
$(x,y_{\textup{high}})$, where $1\leq y_{\textup{low}}<y_{\textup{high}}$,
the distance between $(1,1)$ and $(x,y_{\textup{low}})$ is strictly
smaller than the distance between $(1,1)$ and $(x,y_{\textup{high}})$.
We prove the claim by considering two cases:

\textbf{Case (i)}. Assume that $x$ is odd, and hence both $(x,y_{\textup{low}})$
and $(x,y_{\textup{high}})$ are in $V_{\textup{odd}}$. Consider
a shortest path from $(1,1)$ to $(x,y_{\textup{high}})$. The path
must use a vertex $(x_{\textup{pass}},y_{\textup{low}})\in V_{\textup{odd}}$
at $y$-coordinate $y_{\textup{low}}$. From $(x_{\textup{pass}},y_{\textup{low}})$
we can reach $(x,y_{\textup{low}})$ in exactly $|x-x_{\textup{pass}}|/2$
steps. The number of steps required to reach $(x,y_{\textup{high}})$
from $(x_{\textup{pass}},y_{\textup{low}})$ is strictly greater than
$|x-x_{\textup{pass}}|/2$ since some steps must be used to increase
the $y$-coordinate so it will eventually reach $y_{\textup{high}}$,
and for each such up-move the $x$-coordinate is increased/decreased
only by $1$. Thus, if $x$ is odd then the distance between $(1,1)$
and $(x,y_{\textup{low}})$ is strictly smaller than the distance
between $(x,y_{\textup{high}})$.

\textbf{Case (ii)}. Assume that $x$ is even, and hence both $(x,y_{\textup{low}})$
and $(x,y_{\textup{high}})$ are in $V_{\textup{even}}$. We will
use the same argument as for odd values of $x$, only with the difference
that we consider a vertex $(x_{\textup{pass}},y_{\textup{low}}-1)\in V_{\textup{odd}}$
on a shortest path from $(1,1)$ to $(x,y_{\textup{high}})$. From
$(x_{\textup{pass}},y_{\textup{low}}-1)$ we can reach $(x,y_{\textup{low}})$
in at most $\left\lfloor |x-x_{\textup{pass}}|/2\right\rfloor +1$
steps, where the $+1$ comes from the fact that we need to go up one
$y$-coordinate. The number of steps required to reach $(x,y_{\textup{high}})$
from $(x_{\textup{pass}},y_{\textup{low}}-1)$ is strictly greater
than $\left\lfloor |x-x_{\textup{pass}}|/2\right\rfloor +1$ since
some steps must be used to increase the $y$-coordinate so it will
eventually reach $y_{\textup{high}}$, and for each such up-move the
$x$-coordinate is increased/decreased only by $1$. Thus, also for
even values of $x$ we have that the distance between $(1,1)$ and
$(x,y_{\textup{low}})$ is strictly smaller than the distance between
$(1,1)$ and $(x,y_{\textup{high}})$.

\textbf{Claim 2}. For any two vertices $(x,y_{\textup{low}})$ and
$(x,y_{\textup{high}})$, where $y_{\textup{low}}<y_{\textup{high}}\leq1$,
the distance between $(1,1)$ and $(x,y_{\textup{high}})$ is strictly
smaller than the distance between $(1,1)$ and $(x,y_{\textup{low}})$.
We prove the claim by using exactly the same reasoning as for Claim~1.

Using Claim~1 and~2 we conclude that there are at most two vertices
$(x,y)$ and $(x,y')$, with the same $x$-coordinate, at distance
$d$ from $(1,1)$. The leftmost vertex that is at distance $d$ from
from $(1,1)$ is $(1-2d,1)$. It is reached by making $d$ consecutive
left-moves. Similarly, the rightmost vertex at distance $d$ from
$(1,1)$ is $(1+2d,1)$. Thus, the $x$-coordinate of any vertex at
distance $d$ from $(1,1)$ is in the set $\{1-2d,\dots,1+2d\}$,
and hence there are at most $2\times(4d+1)=8d+2$ vertices at distance
$d$ from $(1,1)$. That is, $n_{d}((1,1))\leq8d+2$. We have now
showed that for any vertex $v\in V_{\textup{odd}}$,
$\left\lfloor d/2\right\rfloor \leq n_{d}(v)\leq8d+2$.

It remains to derive upper and lower bounds on $n_{d}(v)$ for $v\in V_{\textup{even}}$.
Without loss of generality we assume that $v=(0,0)$ is the vertex
at $x$-coordinate $0$ and $y$-coordinate $0$. Fix any positive
integer $d$.

We derive a lower bound on $n_{d}((0,0))$ in the same way as when
$v=(1,1)$. For each odd value of $y\in\{1,\dots,d\}$, let $(x,y)$
be the vertex at distance $d$ from $(0,0)$ that is reached with
the following path: $(0,0),(1,1),(2,2),\dots,(y,y),(y+2,y),(y+4,y)\dots,(x,y)$.
Thus, there are at least $\left\lfloor d/2\right\rfloor $ vertices
at distance $d$ from $(0,0)$, and we have
$n_{d}((0,0))\geq\left\lfloor d/2\right\rfloor $.

We now derive an upper bound on $n_{d}((0,0))$. Vertex $(0,0)$ has
exactly four neighbours: $(1,1)$, $(1,-1)$, $(-1,-1)$ and $(-1,1)$,
which are all in $V_{\textup{odd}}$. The shortest path from $(0,0)$
to any vertex at distance $d$ from $(0,0)$ must use one of these
four vertices. Thus, an upper bound on the number of vertices at distance
$d$ from $(0,0)$ is $n_{d}((0,0))\leq
n_{d-1}((1,1))+n_{d-1}((1,-1))+n_{d-1}((-1,-1))+n_{d-1}((-1,1))$.
From the upper bound above we have that there are at most $8(d-1)+2$
vertices at distance $d-1$ from a vertex in $V_{\textup{odd}}$.
Hence there are at most than $4\times(8(d-1)+2)=32d-24$ vertices
at distance $d$ from $(0,0)$, and we have $n_{d}((0,0))\leq32d-24$.

Finally, for any vertex $v\in V_{\mathcal{G}}$ and any positive integer
$d$ we have shown that $\left\lfloor d/2\right\rfloor \leq n_{d}(v)\leq32d-24$.
\end{proof}

For a vertex $v\in V_{\mathcal{G}}$ and an integer $d\geq0$, let
$\ball_{d}(v)$ denote the set of vertices that are at most distance
$d$ from $v$. Thus we have $\ball_{0}(v)=\{v\}$.

\begin{lemma}
\label{lem:ratio}For any real number $a>0$ there is an integer $d\geq0$
such that \textup{\[
\frac{|\partial\ball_{d}(v)|}{|\ball_{d}(v)|}\leq a,\]
uniformly in $v\in V_{\mathcal{G}}$.}
\end{lemma}
\begin{proof}
Let $v$ be a vertex in $V_{\mathcal{G}}$ and let $a > 0$ be a real number.
For an integer $d \geq 0$,
let $n_{d}(v)$ denote the number of vertices at distance
$d$ from $v$. By Lemma~\ref{lem:sub-exponential}, $n_{d}(v)\in\Theta(d)$.
We have $|\partial\ball_{d}(v)|=n_{d+1}(v)\in\Theta(d)$ and
$|\ball_{d}(v)|=\sum_{i=0}^{d}n_{d}(v)\in\Theta(d^{2})$.
Hence there is an integer $d_0 \geq 0$ such that
$|\partial\ball_{d}(v)|/|\ball_{d}(v)| \leq a$
for $d \geq d_0$.
\end{proof}

\subsection{The Markov chain $\mathcal{M}_{d}$}

In order to analyse the mixing time of the Glauber dynamics we first
define a similar Markov chain that corresponds to heat-bath dynamics
on small subregions instead of single vertices. For a region $R$,
vertex $v\in V_{\mathcal{G}}$ and integer $d\geq0$, let $R_{v}^{d}=R\cap\ball_{d}(v)$.
Let $R^{d}=\{v\in V_{\mathcal{G}}\,|\, R_{v}^{d}\neq\emptyset\}.$
For a region $R$, $q_{0}$-colouring $\mathcal{B}$ of $\partial R$
and integer $d\geq0$, we define the heat-bath Markov chain $\mathcal{M}_{d}$
as follows. The state space is $\Omega_{R}(\mathcal{B})$ and a transition
from a state $\sigma$ is made in the following way: First choose
a vertex $v$ uniformly at random from $R^{d}$. Let $\mathcal{B}_{v}^{d}$
be the colouring of $\partial R_{v}^{d}$ induced by $\sigma$ and
$\mathcal{B}$. To make the transition from $\sigma$, recolour the
vertices in $R_{v}^{d}$ by sampling a colouring from $\pi_{\mathcal{B}_{v}^{d}}$,
the uniform distribution on proper colourings of the region $R_{v}^{d}$
that agree with $\partial R_{v}^{d}$. As for the Glauber dynamics,
the stationary distribution of $\mathcal{M}_{d}$ is $\pi_{\mathcal{B}}$.
Since $\ball_{0}(v)=\{v\}$, Glauber dynamics is $\mathcal{M}_{0}$.
In order to prove rapid mixing of the Glauber dynamics,
we will use the mixing time of $\mathcal{M}_d$ for some constant $d$ and use a
Markov chain comparison method to infer rapid mixing
of $\mathcal{M}_{0}$.

To establish the mixing time of $\mathcal{M}_d$ we use path coupling, due to
Bubley and Dyer~\cite{bd-pctprmmc-97}. Let $\sigma_1$ and $\sigma_2$ be two states of
$\mathcal{M}_d$, where $d$ is to be specified. Using the path-coupling method,
we only need to consider two colourings $\sigma_1$ and $\sigma_2$ that differ on exactly
one vertex, which we refer to as $w$. That is, the Hamming distance between
$\sigma_1$ and $\sigma_2$ is~1. Let $\mathcal{M}_d$ make a transition from
$\sigma_1$ to $\sigma'_1$, and from $\sigma_2$ to $\sigma'_2$.
We want to correlate (or couple) these two transitions such that the expected
Hamming distance between $\sigma'_1$ and $\sigma'_2$ is less than~1.
If we can do this then we use the path-coupling theorem
(see for instance~\cite{bd-pctprmmc-97,dg-rwco-99}]) to infer the mixing time of
$\mathcal{M}_d$. It is possible to construct such a coupling of the transitions
provided $d$ is sufficiently large. The idea is that we update the same vertices
$R_{v}^{d}$ in both the transition from $\sigma_1$ to $\sigma'_1$ and
$\sigma_2$ to $\sigma'_2$. If the vertices we update do not include
$w$, and $w$ is not in $\partial R_{v}^{d}$, then we choose the same colouring of
$R_{v}^{d}$ in both transitions, and hence the Hamming distance between
$\sigma'_1$ and $\sigma'_2$ remains~1. If the vertices $R_{v}^{d}$
we update contain $w$ then again we choose the same colouring of $R_{v}^{d}$ in both
transitions, and the Hamming distance drops to~0. The only situation when the
Hamming distance can increase is when $w$ is on the boundary $\partial R_{v}^{d}$ of the
vertices $R_{v}^{d}$ we update. In this case we use the coupling in
Lemma~\ref{lem:edges-to-vertices} to colour the vertices in $R_{v}^{d}$.
This guarantees that the expected Hamming distance between
$\sigma'_1$ and $\sigma'_2$ will only increase by at most a constant
$K = 30/(\epsilon (1 - \epsilon))$.
Due to Lemma~\ref{lem:ratio} we can choose a radius $d$ such that the ratio of the
probability of having
$w \in \partial R_{v}^{d}$
and the probability of having
$w \in R_{v}^{d}$ is arbitrarily small. Thus, we choose $d$ such that the probability
of decreasing the Hamming distance by~1 is so much bigger than the probability of
increasing it by $K$ that the expected Hamming distance between $\sigma'_1$ and
$\sigma'_2$ is less than~1. The exact details of how to achieve this is explained
in Sections~7.1 and~7.2 in~\cite{gmp-ssmfc-04}.
In Section~7.2 in~\cite{gmp-ssmfc-04}
a proof of the following lemma is found. Note that the notation
in~\cite{gmp-ssmfc-04} differ slightly and of course we make use of
Lemmas~\ref{lem:edges-to-vertices} and~\ref{lem:ratio} as explained above rather than
using equivalent lemmas in~\cite{gmp-ssmfc-04}.

\begin{lemma}
\label{lem:mixing-time-md}Suppose $q=5$. There is an integer $d\geq0$
such that the Markov chain $\mathcal{M}_{d}$ is rapidly mixing on
any region $R$ under any $q_{0}$-colouring $\mathcal{B}$ of $\partial R$.
The mixing time $\tau_{\mathcal{M}_{d}}(\delta)\in O(n\log\frac{n}{\delta})$,
where $n$ is the number of vertices in $R$.
\end{lemma}

\subsection{Rapidly mixing Glauber dynamics}

\label{sec:rapidly-mixing-glauber}We will compare the mixing time
of the Markov chain $\mathcal{M}_{d}$ and the Glauber dynamics $\mathcal{M}_{0}$
by using a method of Diaconis and Saloff-Coste~\cite{ds-c-ctrmv-93}.
Their method has been used before by Goldberg, Martin and Paterson
in~\cite{gmp-ssmfc-04} to compare the mixing time of $\mathcal{M}_{d}$
and $\mathcal{M}_{0}$ under the assumption that $q\geq\Delta+2$,
where $\Delta$ is the maximum degree of the lattice. Here we consider
$q=5$ on the kagome lattice ($\Delta=4$) and therefore we cannot
make direct use of the comparison in~\cite{gmp-ssmfc-04}. Next we
review the comparison described in~\cite{gmp-ssmfc-04} and provide
a proof of rapidly mixing Glauber dynamics with $q=5$ colours. For
a survey on Markov chain comparison in general, see~\cite{dgjm-mcc-06}.

Let $P_{d}$ and $P_{0}$ denote the transition matrix for the chain
$\mathcal{M}_{d}$ and $\mathcal{M}_{0}$, respectively. For $i\in\{0,d\}$,
let $E_{i}$ be the set of pairs of distinct colourings $(\sigma_{1},\sigma_{2})$
with $P_{i}(\sigma_{1},\sigma_{2})>0$. The set $E_{i}$ can be thought
of as containing the edges of the transition graph of $\mathcal{M}_{i}$,
and hence we sometimes refer to a pair in $E_{i}$ as an edge. For
every edge $(\sigma_{1},\sigma_{2})\in E_{d}$, let $\mathcal{P}_{\sigma_{1},\sigma_{2}}$
be the set of paths from $\sigma_{1}$ to $\sigma_{2}$ using transitions
of $\mathcal{M}_{0}$. More formally, let $\mathcal{P}_{\sigma_{1},\sigma_{2}}$
be the set of paths $\gamma=(\sigma_{1}=\theta_{0},\theta_{1},
\dots,\theta_{k}=\sigma_{2})$
such that

\begin{itemize}
\item [(1)]each $(\theta_{i},\theta_{i+1})$ is in $E_{0},$ and
\item [(2)]each edge in $E_{0}$ appears at most once on $\gamma$.
\end{itemize}
We write $|\gamma|$ to denote the length of path $\gamma$. So, for
example, if $\gamma=(\theta_{0},\dots,\theta_{k})$ we have $|\gamma|=k$.
Let $\mathcal{P}=\cup_{(\sigma_{1},\sigma_{2})\in
E_{d}}\mathcal{P}_{\sigma_{1},\sigma_{2}}$
be the set of all paths for all edges in $E_{d}$.

A \emph{flow} is a function $\phi$ from $\mathcal{P}$ to the interval
$[0,1]$ such that for every $(\sigma_{1},\sigma_{2})\in E_{d}$,\[
\sum_{\gamma\in\mathcal{P}_{\sigma_{1},\sigma_{2}}}\phi(\gamma)=
P_{d}(\sigma_{1},\sigma_{2})\pi_{\mathcal{B}}(\sigma_{1}).\]
For every $(\theta_{1},\theta_{2})\in E_{0}$, the \emph{congestion}
\emph{of edge} $(\theta_{1},\theta_{2})$ in the flow $\phi$ is the
quantity\[
A_{\theta_{1},\theta_{2}}(\phi)=\frac{1}{\pi_{\mathcal{B}}(\theta_{1})P_{0}
(\theta_{1},\theta_{2})}\sum_{\gamma\in\mathcal{P}:(\theta_{1},\theta_{2})\in
\gamma}|\gamma|\phi(\gamma).\]
The \emph{congestion of the flow} is the quantity\[
A(\phi)=\max_{(\theta_{1},\theta_{2})\in E_{0}}A_{\theta_{1},\theta_{2}}(\phi).\]

Theorem~\ref{thm:mixing-times-md-m0-related} below describes how
the mixing times of $\mathcal{M}_{d}$ and $\mathcal{M}_{0}$ are
related. A proof of this theorem can be found in~\cite[Observation~13]{dgjm-mcc-06}.
As pointed out in~\cite{gmp-ssmfc-04}, this theorem is similar to
Proposition~4 of Randall and Tetali~\cite{rt-agdcmc-00} except
that~\cite[Proposition~4]{rt-agdcmc-00} requires the eigenvalues
of transition matrices to be non-negative. Both results are based
closely on the ideas of Aldous~\cite{a-rwfgrmmc-83}, Diaconis and
Stroock~\cite{ds-gbemc-91}, and Sinclair~\cite{s-ibmr-92}. Let
$\tau_{\mathcal{M}_{d}}(\delta)$ be the mixing time of $\mathcal{M}_{d}$
and let $\tau_{\mathcal{M}_{0}}(\delta)$ be the mixing time of the
Glauber dynamics $\mathcal{M}_{0}$.

\begin{theorem}
\label{thm:mixing-times-md-m0-related}Suppose that $\phi$ is a flow.
Let $p=\min_{\theta\in\Omega_{R}(\mathcal{B})}P_{0}(\theta,\theta)$
and assume that $p>0$. Then for any $0<\delta'<\frac{1}{2}$\[
\tau_{\mathcal{M}_{0}}(\delta)\leq\ln\frac{1}{\delta\cdot\pi_{\textup{min}}}\cdot
\max\left[A(\phi)\left(
\frac{\tau_{\mathcal{M}_{d}}(\delta')}{\ln
\frac{1}{2\delta'}}+1\right),\;\frac{1}{2p}\right]\]
where $\pi_{\textup{min}}=
\min_{\sigma\in\Omega_{R}(\mathcal{B})}\pi_{\mathcal{B}}(\sigma)$.
\end{theorem}

\begin{lemma}
\label{lem:mixing-time-m0-given-a}Suppose that there is a flow $\phi$
such that the congestion $A(\phi)\in O(1)$. Then the mixing time
of the Glauber dynamics $\mathcal{M}_{0}$ on a region $R$ is
$\tau_{\mathcal{M}_{0}}(\delta)\in O(n(n+\log\frac{1}{\delta}))$,
where $n$ is the number of vertices in $R$.
\end{lemma}
\begin{proof}
From Definition~\ref{def:glauber}
of Glauber dynamics,
$p=\min_{\theta\in\Omega_{R}(\mathcal{B})}P_{0}(\theta,\theta)\geq1/q$.
Suppose $\delta'=1/n$.
Then by Lemma~\ref{lem:mixing-time-md} we have
$\tau_{\mathcal{M}_{d}}(\delta')\in O(n\log n)$.
With $A(\phi)\in O(1)$, Theorem~\ref{thm:mixing-times-md-m0-related}
gives\[
\tau_{\mathcal{M}_{0}}(\delta)\leq
\ln\frac{1}{\delta\cdot\pi_{\textup{min}}}\cdot O(1)\cdot
O(n)=O(n(n+\log\frac{1}{\delta})\]
since $\pi_{\textup{min}}\geq1/q^{n}$ and hence $\ln(1/\pi_{\textup{min}})\in O(n)$.
\end{proof}

In order to establish the mixing time of the Glauber dynamics $\mathcal{M}_{0}$
by applying Lemma~\ref{lem:mixing-time-m0-given-a} we have to construct
a flow $\phi$ such that the congestion $A(\phi)\in O(1)$.
Given a $q$-colouring $\sigma$ of a region $R$ and a $q_0$-colouring $\mathcal{B}$ of
$\partial R$,
a \emph{single-vertex update} of a vertex $v \in R$ is a recolouring of $v$ to a
colour $c \in Q$
such that no neighbour of $v$ has colour $c$ in either $\sigma$ or $\mathcal{B}$.
Suppose $R$ is a region and $\sigma_{1}$
and $\sigma_{2}$ are two proper 5-colourings of $R$ that differ on $m$ vertices.
The next two lemmas tell us how a series of $O(m)$ single-vertex updates applied to
$\sigma_1$ can transform $\sigma_1$ to $\sigma_2$.
This sequence of single-vertex updates will be used when constructing the flow $\phi$.

\begin{lemma}
\label{lem:kagome-two-neighbours-equal}Consider the region in
Figure~\ref{fig:kagome-two-neighbours-equal}(a).%
\begin{figure}[tp]
\centering
(a)\hspace{-2mm}\includegraphics[scale=\myscaling]{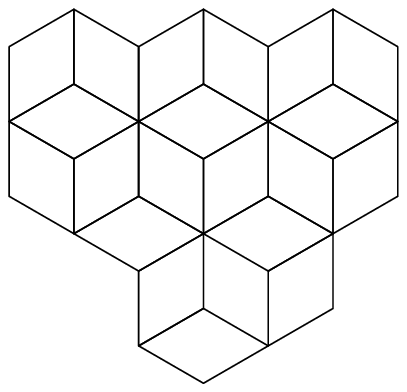}~~~~~~
(b)\hspace{-2mm}\includegraphics[scale=\myscaling]{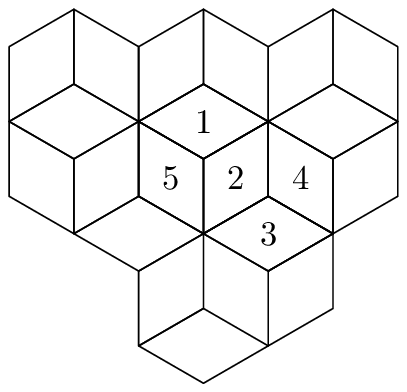}~~~~~~
(c)\hspace{-2mm}\includegraphics[scale=\myscaling]{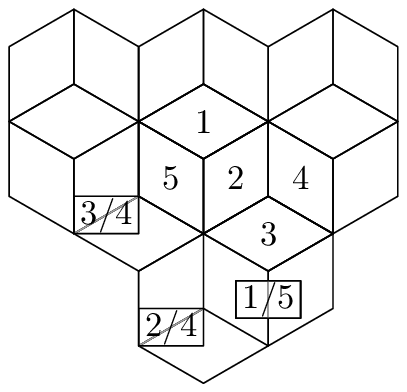}

\medskip

\medskip

\bigskip

\bigskip

(d)\hspace{-2mm}\includegraphics[scale=\myscaling]{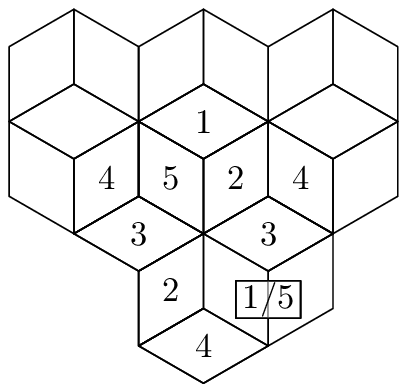}~~~~~~
(e)\hspace{-2mm}\includegraphics[scale=\myscaling]{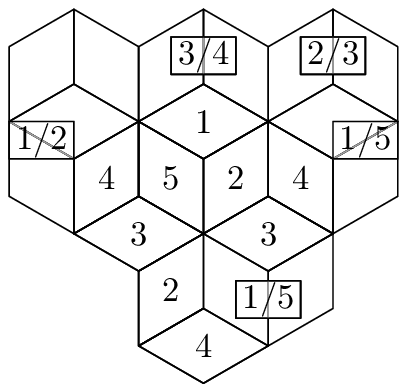}~~~~~~
(f)\hspace{-2mm}\includegraphics[scale=\myscaling]{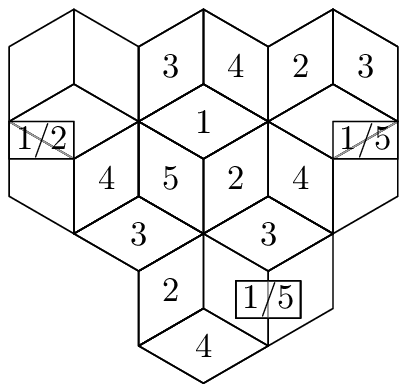}

\medskip

\caption{\label{fig:kagome-two-neighbours-equal}In every proper $5$-colouring
of the region in (a) there is a vertex that has two neighbours with
the same colour.}

\end{figure}
In every proper 5-colouring of this region there is a vertex that
has two neighbours with the same colour.
\end{lemma}
\begin{proof}
Suppose $\sigma$ is a proper 5-colouring of the region in
Figure~\ref{fig:kagome-two-neighbours-equal}(a)
such that no two neighbours of a vertex in the region have the same colour.
We will show that this leads to contradiction.
Without
loss of generality we may assume that five of the vertices
have the colours specified in
Figure~\ref{fig:kagome-two-neighbours-equal}(b).
A vertex is labelled with its colour.
It follows that the two vertices
adjacent to the vertex coloured 5 must have colour 3 and 4, otherwise
there would be a vertex that has two neighbours with the same colour.
Similarly, the vertices adjacent to the vertex coloured 3 must have
colour 1 and 5, and therefore the two bottom left vertices must have
colour 2 and 4 in $\sigma$.
Figure~\ref{fig:kagome-two-neighbours-equal}(c)
illustrates this fact, where a square contains the two colours
of the two vertices it is overlapping. From the
two left squares we see that the colour 4 must be on
the vertices that are as far apart as possible.
Thus, $\sigma$ must agree with
the colouring in Figure~\ref{fig:kagome-two-neighbours-equal}(d).
Figure~\ref{fig:kagome-two-neighbours-equal}(e) illustrates how
other vertices of the region must be coloured in $\sigma$, and
Figure~\ref{fig:kagome-two-neighbours-equal}(f) shows the necessary colouring
of the four rightmost vertices at the top.
To finish the proof
we note that it is impossible to assign colours to the two leftmost vertices
at the top without introducing a vertex such that two of its neighbours
receive the same colour.
\end{proof}

\begin{lemma}
\label{lem:kagome-single-vertex-updates}
Let $R$ be a region
of the kagome lattice and let $\mathcal{B}$ be the 0-colouring of the boundary
$\partial R$.
Suppose that $q = 5$ and let $\sigma_{1}$ and $\sigma_{2}$ be any
two proper $q$-colourings of $R$ that differ on $m$ vertices. We can go from
$\sigma_1$ to $\sigma_2$
by applying a series of $O(m)$ single-vertex updates.
\end{lemma}

\begin{proof}
Let $v \in R$ be a vertex on which $\sigma_{1}$ and $\sigma_{2}$ differ.
We will show how to recolour $v$ to the colour it has in $\sigma_{2}$ by doing at
most a constant number of single-vertex updates. A vertex in $R$ that has the same
colour in both $\sigma_1$ and $\sigma_2$ will not change colour after $v$ has
been updated.
First we analyse situations where no boundary vertices in $\partial R$ are involved.
We note at the end of the proof that if boundary vertices are present,
then it only makes it easier to recolour $v$. That is, assume for now that all vertices
we consider belong to the region $R$. The proof goes through a series of cases.

If possible, simply recolour $v$ to the colour it has in $\sigma_2$.
If this is not possible then there
must be one or two neighbours of $v$ that have
colour $\sigma_{2}(v)$ in $\sigma_{1}$. It cannot be more than two such
neighbours since $\sigma_{1}$ is a proper colouring.

Without loss of generality, assume that $\sigma_{1}(v) = 1$
and $\sigma_{2}(v) = 2$. If $v$ has two neighbours with colour~2 in $\sigma_{1}$
then we will first recolour one of these two neighbours
to some other colour than~2. Let $w$ be the neighbour of $v$ with colour~2 that
we are going to recolour. Note that $\sigma_2(w) \neq 2$ since $\sigma_2$ is a
proper colouring.
If possible, recolour $w$ to some other colour than~2.
If this is not possible then $w$ is ``locked'' and must have
three neighbours coloured~3, 4 and~5, respectively. In this case,
first recolour $v$ (which is possible since $v$ has two neighbours
with colour~2) and then recolour $w$ to colour~1. Now only one
neighbour of $v$ has colour~2. We deal with this case next.

Without loss of generality, assume that $\sigma_{1}(v) = 1$
and $\sigma_{2}(v) = 2$,
and exactly one neighbour $w$ of $v$ has colour~2 in $\sigma_{1}$.
Note that $\sigma_2(w) \neq 2$ since $\sigma_2$ is a proper colouring.
If possible,
recolour $w$ to something else than~2 and then recolour $v$ to~2.
If this is not possible then $w$ is ``locked'' and must have
four neighbours (including $v$) with colours~1,
3, 4 and~5, respectively, in $\sigma_{1}$. Without loss of generality, consider the
region in Figure~\ref{fig:single-site-updates}(a),
\begin{figure}[tp]
\centering
(a)\hspace{-1mm}\includegraphics[scale=\myscaling]{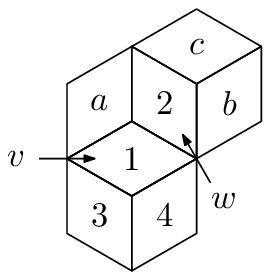}
~~(b)\hspace{-1mm}\includegraphics[scale=\myscaling]{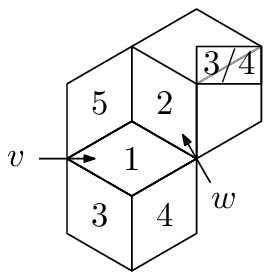}
~~(c)\hspace{-7mm}\includegraphics[scale=\myscaling]{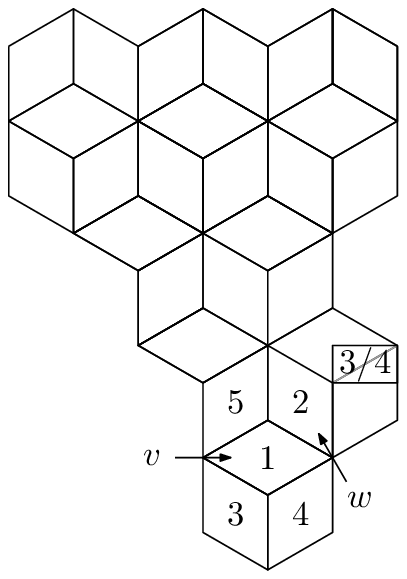}
~~~~~(d)\hspace{-7mm}\includegraphics[scale=\myscaling]{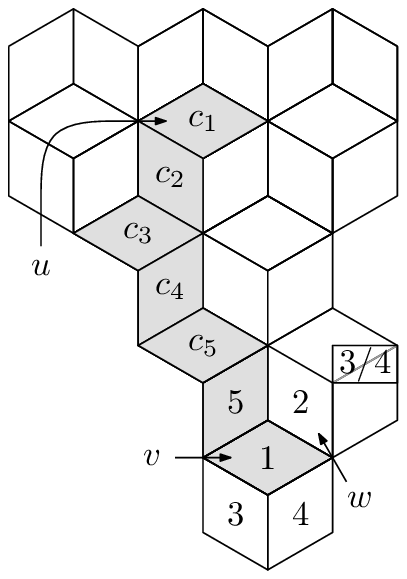}

\smallskip

\caption{\label{fig:single-site-updates}The colours 1 and 2 are going to
swap place. Lemma~\ref{lem:kagome-single-vertex-updates} guarantees
that this can be done with a constant number of single-vertex updates.}

\end{figure}
which is a subregion of $R$. Call this region $R'$.
The vertices of $R'$ are labelled with their colours in $\sigma_1$.
The vertex with colour~1
is $v$ and the vertex with colour~2 is $w$. We assume without loss
of generality that the two neighbours of $v$ that are below $v$ are the
two neighbours with colour~3 and~4 in $\sigma_1$.

Three of the vertices in $R'$ are given the colours
$a$, $b$ and $c$, which are to be determined.
Since $w$ is ``locked'', the colours $a$, $b$ and $c$ is any permutation of the
colours~3, 4 and~5.
If $a$ is~3 or~4
then we recolour $v$ to~5 and then recolour $w$ to~1, and then
recolour $v$ to~2. If this is not the case then $a$ must be~5, and
hence the colours $b$ and $c$ are~3
and~4 in any order.
Figure~\ref{fig:single-site-updates}(b) illustrates this. We now analyse this case.

We will use Lemma~\ref{lem:kagome-two-neighbours-equal}
to show that we can recolour $v$ to~2 without changing the colour
of any other vertex except $w$ (which will be recoloured to~1).
Consider Figure~\ref{fig:single-site-updates}(c) which
illustrates the region $R'$ extended with vertices in $R$. The vertices we
extend $R'$ with correspond to the region that we used in
Lemma~\ref{lem:kagome-two-neighbours-equal}.
From Lemma~\ref{lem:kagome-two-neighbours-equal}
we know that there must be at least one vertex $u$ among the vertices we extend $R'$ with
such that $u$ has two neighbours with the same colour.
Let $P$ be a shortest path from $v$ to $u$ such that the path goes from
$v$ to the neighbour above that has colour~5 and then is entirely inside
the region we added to $R'$. Figure~\ref{fig:single-site-updates}(d) illustrates
an example of such a path. The path is shaded in the figure.
Suppose that the vertex $u$ is chosen such that all vertices on the path $P$
(except from $u$ itself)
are ``locked'' (have four neighbours of different colours). Note that if the vertex coloured~5
above $v$ does not have four neighbours of different colours then we let $u$ be this
vertex and hence the path
$P$ consists only of the two vertices $u$ and $v$.

Suppose that the path $P$ contains
$k$ vertices. Let $c_{1},\dots,c_{k}$ be the colours in $\sigma_{1}$
of the vertices from $u$ to $v$ along the path. That is, $\sigma_{1}(u)=c_{1}$,
$c_{k-1}=5$ and $c_{k}=1$. Since $u$ has two neighbours with the
same colour, we recolour $u$ from $c_{1}$ to another colour~$c_{1}'$.
Now the vertex after $u$ on $P$ has two neighbours with
the same colour (namely $c_{1}'$), since all its neighbours had different colours before
recolouring $u$. We recolour this vertex from $c_{2}$ to $c_{2}'$. We continue
this recolouring procedure along the path $P$ all the way to vertex $v$,
which will be recoloured to~3. Note that the vertex above $v$ which had previously
colour~5 now must have colour~3 or~4.
We can now recolour $w$ to~1 and then recolour $v$ to~2.
It remains to recolour the vertices
on the path back to their original colours in $\sigma_1$. We do this by reversing
the recolouring procedure, starting with the vertex above $v$, which
is recoloured back to~5. When $u$ is recoloured back to $c_{1}$ we are
done.

We have now shown how a constant number of single-vertex updates
are applied in order to recolour a vertex $v$ to the colour it has
in $\sigma_{2}$ without changing the colour on vertices that have the
same colour in $\sigma_1$ and $\sigma_2$.

We note that if any vertices involved in the recolouring procedure
of $v$ are boundary vertices then this will only make it easier.
Note from the statement of the lemma that we assume that a boundary vertex has colour~0.
As we have seen, the tricky
situations arise when a vertex is ``locked'' with four neighbours
of different colours (excluding colour~0). Such a vertex is tricky because we cannot
just change its colour to another colour in $Q = \{1,\dots,5\}$.
A vertex that is adjacent to a boundary
vertex
can never be ``locked'' since there is always at least one colour in $Q$ that it can be
recoloured to.
Thus, although the part of the proof above assumes that all vertices are in $R$, we note
that the presence of
boundary vertices only makes the recolouring procedure easier. Of course, depending on
which vertex $v$
we are going to recolour, and which neighbour $w$ is ``locked'', the path $P$ might go in
a direction that is different from the one in Figure~\ref{fig:single-site-updates}(d).
However, the same
technique is applied in order to successfully recolour~$v$.

Finally, in order to transform $\sigma_{1}$ to $\sigma_{2}$, we recolour each
vertex $v$ at which $\sigma_{1}$ and $\sigma_{2}$ differ. For each
such vertex it takes only a constant number of single-vertex updates
to do so. Since $\sigma_{1}$ and $\sigma_{2}$ differ only at $m$
vertices, the total number or updates is $O(m)$.
Notice that in recolouring a vertex $v$ we might have changed the colours of neighbours
of $v$ as well. However, we never change the colour of a vertex whose colour
agrees with the destination colouring $\sigma_{2}$, a fact that ensures that
the process described above indeed terminates with the colouring $\sigma_{2}$.
\end{proof}

We are now able to show how to construct a flow $\phi$ such that
$A(\phi)\in O(1)$ for $q=5$ colours. This only holds when the boundary
colouring $\mathcal{B}$ of $\partial R$ is the 0-colouring.

\begin{lemma}
\label{lem:constant-congestion}Suppose $q=5$. Consider any region
$R$ and let $\mathcal{B}$ the the 0-colouring of $\partial R$.
There is a flow $\phi$ such that the congestion $A(\phi)\in O(1)$.
\end{lemma}
\begin{proof}
For every pair $(\sigma_{1},\sigma_{2})\in E_{d}$ we know that $\sigma_{1}$
and $\sigma_{2}$ differ only on vertices that are contained in the
ball $\ball_{d}(v)$ for some vertex $v\in R^{d}$. Let $\prec$ be
a fixed canonical ordering of the vertices in $R$.
Let $\gamma_{\sigma_{1},\sigma_{2}}\in\mathcal{P}_{\sigma_{1},\sigma_{2}}$
be the path from $\sigma_{1}$ to $\sigma_{2}$ constructed according
to the proof of Lemma~\ref{lem:kagome-single-vertex-updates}. We
consider vertices in order specified by~$\prec$ to make sure that
$\gamma_{\sigma_{1},\sigma_{2}}$ is well defined.

Assign all of the flow from $\sigma_{1}$ to $\sigma_{2}$ to path
$\gamma_{\sigma_{1},\sigma_{2}} \in \mathcal{P}_{\sigma_{1},\sigma_{2}}$.
That is,
$\phi(\gamma_{\sigma_{1},\sigma_{2}})=
P_{d}(\sigma_{1},\sigma_{2})\pi_{\mathcal{B}}(\sigma_{1})$
and $\phi(\gamma)=0$ for all paths
$\gamma\in\mathcal{P}_{\sigma_{1},\sigma_{2}}
\backslash\{\gamma_{\sigma_{1},\sigma_{2}}\}$.
Let $\theta_{1}$ and $\theta_{2}$, where $(\theta_{1},\theta_{2})\in E_{0}$,
be two colourings that disagree on a vertex $w$. Then the congestion
of edge $(\theta_{1},\theta_{2})$ is
\begin{eqnarray*}
A_{\theta_{1},\theta_{2}}(\phi) & = &
\frac{1}{\pi_{\mathcal{B}}(\theta_{1})P_{0}(\theta_{1},\theta_{2})}
\sum_{\substack{(\sigma_{1},\sigma_{2})\in E_{d}:\\
(\theta_{1},\theta_{2})\in\gamma_{\sigma_{1},\sigma_{2}}}
}|\gamma_{\sigma_{1},\sigma_{2}}|P_{d}(\sigma_{1},\sigma_{2})
\pi_{\mathcal{B}}(\sigma_{1})\\
& = & \sum_{\substack{(\sigma_{1},\sigma_{2})\in E_{d}:\\
(\theta_{1},\theta_{2})\in\gamma_{\sigma_{1},\sigma_{2}}}
}|\gamma_{\sigma_{1},\sigma_{2}}|\cdot\frac{P_{d}(\sigma_{1},
\sigma_{2})}{P_{0}(\theta_{1},\theta_{2})}\cdot
\frac{\pi_{\mathcal{B}}(\sigma_{1})}{\pi_{\mathcal{B}}(\theta_{1})}\\
 & \leq & \sum_{\substack{(\sigma_{1},\sigma_{2})\in E_{d}:\\
(\theta_{1},\theta_{2})\in\gamma_{\sigma_{1},\sigma_{2}}}
}k_{1}\cdot\frac{P_{d}(\sigma_{1},
\sigma_{2})}{P_{0}(\theta_{1},\theta_{2})}\,\,\leq\,\,
k_{1}\cdot k_{2}\,\,\leq\,\, O(1),
\end{eqnarray*}
where $k_{1}$ and $k_{2}$ are constants, specified next.
Note that $\pi_{\mathcal{B}}(\sigma_{1})/\pi_{\mathcal{B}}(\theta_{1})=1$.

The path length $|\gamma_{\sigma_{1},\sigma_{2}}|$ is upper-bounded
by a constant $k_{1}$ since $\sigma_{1}$ and $\sigma_{2}$ differ
only on vertices inside a ball of fixed radius $d$.
The path $\gamma_{\sigma_{1},\sigma_{2}}$
is constructed such that for each vertex $v$ that is updated, we
do at most a constant number of recolourings of vertices that are
within constant distance from $v$.

To see that the last sum is bounded by a constant $k_{2}$, note that
there are only a constant number of pairs $(\sigma_{1},\sigma_{2})$
in the summation. This is true since $\sigma_{1}$ and $\sigma_{2}$
agree with $\theta_{1}$ on all vertices in $R$ except in a constant-sized
ball around a vertex $w$ on which $\theta_{1}$ and $\theta_{2}$
differ. Let $m$ be the number of vertices $u$ such that $R_{u}^{d}$
contains all vertices on which $\sigma_{1}$ and $\sigma_{2}$ differ.
Note that $m$ is bounded by a constant since $\sigma_{1}$ and $\sigma_{2}$
differ only on vertices inside a ball of fixed radius $d$. We have
\[
P_{d}(\sigma_{1},\sigma_{2})\leq\frac{m}{|R^{d}|}\in O(\frac{1}{|R^{d}|}).\]
 Furthermore,\[
P_{0}(\theta_{1},\theta_{2})\geq
\frac{1}{|R^{d}|}\cdot\frac{1}{q}\in\Omega(\frac{1}{|R^{d}|})\]
since $1/q$ is the smallest probability of making a transition in
$\mathcal{M}_{0}$ from colouring $\theta_{1}$ to $\theta_{2}$ once
vertex $w$ on which $\theta_{1}$ and $\theta_{2}$ differ has been
chosen for an update. Thus,\[
\frac{P_{d}(\sigma_{1},\sigma_{2})}{P_{0}(\theta_{1},\theta_{2})}\in O(1)\]
and we have that the sum is bounded by a constant $k_{2}$.

Now, $A_{\theta_{1},\theta_{2}}(\phi)\in O(1)$ for all
$(\theta_{1},\theta_{2})\in E_{0}$
and it follows that the congestion $A(\phi)\in O(1)$.
\end{proof}
Finally we have the machinery for proving Theorem~\ref{thm:kagome-glauber}.

\begin{theorem*}
[\ref{thm:kagome-glauber}, repeated]For any region $R$ of the kagome
lattice and $q=5$ colours, the Glauber dynamics is rapidly mixing
on $R$ under the 0-colouring of $\partial R$.
The mixing time $\tau(\delta)\in O(n^{2}+n\log\frac{1}{\delta})$,
where $n$ is the number of vertices in $R$.
\end{theorem*}
\begin{proof}
The theorem is proved by using Lemmas~\ref{lem:mixing-time-m0-given-a}
and~\ref{lem:constant-congestion}.
\end{proof}

The proof of Theorem~\ref{thm:edge-dynamics}
is similar to the proof of
Theorem~\ref{thm:kagome-glauber}. The implications from rapid mixing of
$\mathcal{M}_d$ to rapid mixing of the heat-bath dynamics on edges hold.
Lemma~\ref{lem:mixing-time-m0-given-a} has to be stated with
$\mathcal{M}_0$ replaced by the heat-bath dynamics on edges
(which slightly changes the proof) and
Lemma~\ref{lem:constant-congestion} has be adjusted to deal with
an arbitrary $q_0$-colouring of the boundary of the region, where $q = 5$.
Showing that the congestion is constant under any
$q_0$-colouring of the boundary
is not difficult
since we are allowed to update two vertices at the same time.


\section{The computational part of Lemma~\ref{lem:gamma-exp-decay-final}}
\label{sec:computational-part}

The computational part of the proof
of Lemma~\ref{lem:gamma-exp-decay-final}
consists of two tasks:
calculating the values $\mu_{F_{f,m}}$
and constructing
an $(\calA,\calF)$-set $\calS$ that is good with respect to
$\epsilon = 1/1000$.
These two tasks are explained in the next sections.
Both tasks are carried out using computer assistance.
We have written programs in C, and the source code can be found on the webpage
\myurl

\subsection{Computing $\mu_{f,m}$}

Calculating the values $\mu_{F_{f,m}}$ is a computationally challenging task.
We are going to to calculate $\mu_{F_{f,m}}$ for
$f \in \{1,\dots,4720\}$ and $m \in \{1,\dots,4\}$.
From the definition of $\mu_{F_{f,m}}$ in Section~\ref{sec:the-constants},
$\mu_{F_{f,m}} = 0$ if $F'_{f,m} = \emptyset$.
For every fixed $f \in \{1,\dots,4720\}$,
$F'_{f,m} = \emptyset$ for exactly two values of $m \in \{1,\dots,4\}$.
Thus, we will have to calculate the value of
$2 \times 4720 = 9440$ constants $\mu_{F_{f,m}}$.
We must be able to compute a single value rather quickly, otherwise
the total running time for all values will be too long. A brute-force
approach would result in a running time of several months, maybe even
years. We use a technique that is illustrated with the following example.

\begin{figure}[tp]
\centering
(a)\hspace{-2mm}\includegraphics[scale=\myscaling]{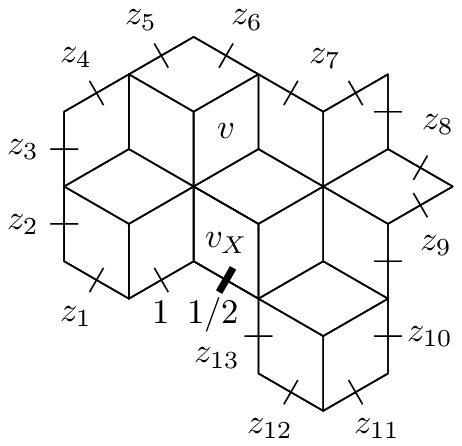}\qquad~~~~
(b)\hspace{-2mm}\includegraphics[scale=\myscaling]{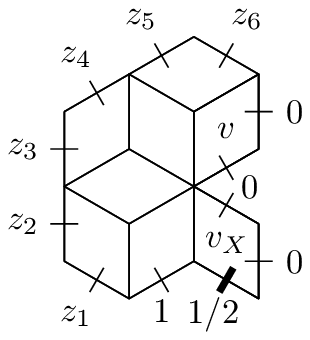}\qquad~~~~
(c)\hspace{-1mm}\includegraphics[scale=\myscaling]{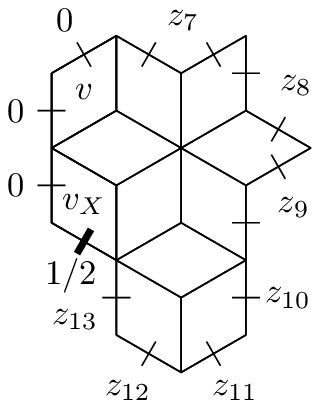}

\smallskip

\caption{\label{fig:kagome-splitting-region}The region $R_{X}$ of the edge-boundary
pair $X$ in (a) is split into two halves (b) and (c). The split is
along the vertices labelled $v$ and $v_{X}$. }

\end{figure}

Suppose $\exR_{F_{f}}$ is the extended region in
Figure~\ref{fig:kagome-splitting-region}(a)
and suppose $m = 1$.
Hence the set $F'_{f,m} \neq \emptyset$.
The value $\mu_{F_{f,m}}$ is obtained
by maximising $\mu_{1,2}(X)$ over all edge-boundary pairs
$X\in F'_{f,m}$.
Let $v'$ be the vertex that is a neighbour to both $v_X$ and $w_X$.
Since $m = 1$,
the boundary edge between $w_X$ and $v'$ has colour~1
in every edge-boundary pair $X\in F'_{f,m}$.
An edge-boundary pair $X\in F'_{f,m}$ is therefore uniquely specified
by the colour of its remaining
boundary edges.
Figure~\ref{fig:kagome-splitting-region}(a)
illustrates an arbitrary edge-boundary pair $X\in F'_{f,m}$,
where boundary edges are labelled
with their colour ($z_1, \dots, z_{13}$).
Thus, in order to maximise $\mu_{1,2}(X)$
over all $X\in F'_{f,m}$, we could
loop through all combinations of the colours $z_{1},\dots,z_{13}$
and compute $\mu_{1,2}(X)$ for each such combination.
This process will take very long.
Next we explain how to speed up the process.

By computing $\mu_{1,2}(X)$ for many colourings of the boundary,
one quickly makes the observation
that only some particular colourings of the boundary result in
a large value of $\mu_{1,2}(X)$. For other colourings,
$\mu_{1,2}(X)$ tends to be rather small.
For example, it turns out that if
$z_{1},\dots,z_{6}$ are all colour
1, then $\mu_{1,2}(X)$ will be rather small regardless
of the remaining colours $z_{7},\dots,z_{13}$.
Thus, setting
the colours $z_{1},\dots,z_{6}$ to~1 is a ``bad'' choice
if we want to maximise $\mu_{1,2}(X)$.
From this observation we conclude that if we can
filter out certain ``bad'' colourings
of the boundary then we could speed up the process of finding the
maximum value $\mu_{1,2}(X)$.

We ``split'' the extended region $\exR_{F_{f}}$ into two extended
regions $\exR^{\textup{left}}$ and $\exR^{\textup{right}}$.
Figure~\ref{fig:kagome-splitting-region}(b)
and~(c) illustrate $\exR^{\textup{left}}$ and $\exR^{\textup{right}}$,
respectively. The two regions share the vertices in the split. In
this case it is vertex $v_{X}$ and $v$, both labelled in the figure.
Let $X^{\textup{left}}$ be the edge-boundary pair such that
$R_{X^{\textup{left}}}=\exR^{\textup{left}}$,
$e_{X^{\textup{left}}}=e_{X}$ and the boundary edges receive the
same colours as in $X$. Boundary edges that are introduced from the
split are given colour~0. Let $X^{\textup{right}}$ be the edge-boundary
pair defined similarly to $X^{\textup{left}}$ but with
$R_{X^{\textup{right}}}=\exR^{\textup{right}}$.
Figure~\ref{fig:kagome-splitting-region}(b) and~(c) illustrate
$X^{\textup{left}}$ and $X^{\textup{right}}$.

Let $B$ be the colouring of $\mathcal{E}R_{X}$ such that $B(e)=B_{X}(e)$
for $e \in \mathcal{E}R_{X} \backslash \{e_{X}\}$ and $B(e_{X}) = 0$.
Recall from Section~\ref{sec:gamma-exp-decay} that for $i \in Q$,
$n_i(X)$ denotes the number of proper $q$-colourings
$\sigma$
in $\Omega_{R_X}(B)$ such that $\sigma(v_X) = i$.
For two colours
$i,i'\in Q$
we now define $n^\textup{both}_{i,i'}$ to be the number of proper $q$-colourings
$\sigma$
in $\Omega_{R_X}(B)$ such that $\sigma(v_X) = i$
and $\sigma(v)=i'$, where $v$ is the second vertex in the split.
Thus,
$$
n_{i}(X)=\sum_{i'=1}^{q}n^\textup{both}_{i,i'}.
$$
Let $B^\textup{left}$ be the colouring of
$\mathcal{E}R_{X^\textup{left}}$ such that
$B^\textup{left}(e_{X^\textup{left}}) = 0$ and
$B^\textup{left}(e)=B_{X^\textup{left}}(e)$
for $e \in \mathcal{E}R_{X^\textup{left}} \backslash
\{e_{X^\textup{left}}\}$.
For two colours
$i,i'\in Q$
we define $n^\textup{left}_{i,i'}$
to be the number of proper $q$-colourings
$\sigma$
in $\Omega_{R_{X^\textup{left}}}(B^\textup{left})$ such that $\sigma(v_X) = i$
and $\sigma(v)=i'$.
We define $n^\textup{right}_{i,i'}$ similarly for the edge-boundary pair
$X^{\textup{right}}$.
It follows that
$$
n_{i,i'}^\textup{both}=n_{i,i'}^{\textup{left}}n_{i,i'}^{\textup{right}},
$$
and hence
$$
n_{i}(X)=\sum_{i'=1}^{q}n_{i,i'}^{\textup{left}}n_{i,i'}^{\textup{right}}.
$$
With $q=5$ colours, we have
\begin{align}
\mu_{1,2}(X)\; & =\;\frac{n_{1}(X)}{\sum_{i\in\{1,3,4,5\}}
n_{i}(X)}\;=\;\frac{\sum_{j=1}^{5}n_{1,j}^{\textup{left}}n_{1,j}^{\textup{right}}}{
\sum_{i\in\{1,3,4,5\}}\sum_{k=1}^{5}n_{i,k}^{\textup{left}}n_{i,k}^{\textup{right}}}
\nonumber \\
 & =\;\sum_{j=1}^{5}\frac{n_{1,j}^{\textup{left}}n_{1,j}^{\textup{right}}}{
\sum_{i\in\{1,3,4,5\}}\sum_{k=1}^{5}n_{i,k}^{\textup{left}}n_{i,k}^{\textup{right}}}
\nonumber \\
&
=\;\sum_{j=1}^{5}\frac{1}{\sum_{i\in\{1,3,4,5\}}
\sum_{k=1}^{5}\left(\frac{n_{i,k}^{\textup{left}}}{n_{1,j}^{\textup{left}}}\times
\frac{n_{i,k}^{\textup{right}}}{n_{1,j}^{\textup{right}}}\right)}.
\label{eq:kagome-dominating-colouring}
\end{align}
Note that the colours $z_{1},\dots,z_{6}$ specify the
quantity $n_{i,i'}^{\textup{left}}$,
and $z_{7},\dots,z_{13}$ specify the
quantity $n_{i,i'}^{\textup{right}}$.
In order to maximise $\mu_{1,2}(X)$ over edge-boundary pairs $X$,
we could consider all combinations of the colours
$z_1,\dots,z_{13}$
and use
Equation~(\ref{eq:kagome-dominating-colouring}).
There are $5^{13} \approx 1.2 \times 10^9$ such combinations,
so considering them all will take very long.
Now,
consider two different sets of the six colours $z_1,\dots,z_{6}$.
For $i,i' \in Q$,
let
$n_{i,i'}^{\textup{left-1}}$ be the value of $n_{i,i'}^{\textup{left}}$
for the first set of colours, and let
$n_{i,i'}^{\textup{left-2}}$ be the value of $n_{i,i'}^{\textup{left}}$
for the second set of colours.
Suppose
\begin{equation}
\label{eq:compare-colourings}
\frac{n_{i,k}^{\textup{left-}1}}{n_{1,j}^{\textup{left-}1}}\leq
\frac{n_{i,k}^{\textup{left-}2}}{n_{1,j}^{\textup{left-}2}}
\end{equation}
for all $i\in\{1,3,4,5\}$, $j\in\{1,\dots,5\}$ and $k\in\{1,\dots,5\}$.
Then we have from Equation~(\ref{eq:kagome-dominating-colouring})
that $\mu_{1,2}(X)$ can only get smaller if we take
$n_{i,i'}^{\textup{left}} = n_{i,i'}^{\textup{left-2}}$ instead of
$n_{i,i'}^{\textup{left}} = n_{i,i'}^{\textup{left-1}}$.
In other words,
there is no point considering the colours specified by the second set of colours
$z_1,\dots,z_6$
when maximising $\mu_{1,2}(X)$. This observation suggests that we
loop through all combinations of colours $z_1,\dots,z_6$
and compare each pair of combinations like in Equation~(\ref{eq:compare-colourings}).
We only keep the sets of colours that cannot be ruled out
in some pairwise comparison
like
the second set above.
This gives us a collection $C^{\textup{left}}$ of colours $z_{1},\dots,z_{6}$
that turns out to be much smaller than the collection of all $5^6 = 15,625$
sets of colours.
Similarly we obtain a collection
$C^{\textup{right}}$ of colours $z_{7},\dots,z_{13}$ for the right
part of the region.
In order to find which colours $z_{1},\dots,z_{13}$ that maximise
$\mu_{1,2}(X)$ we combine $C^{\textup{left}}$ with $C^{\textup{right}}$.
That is, we use Equation~(\ref{eq:kagome-dominating-colouring})
to
compute $\mu_{1,2}(X)$
for each set $z_{1},\dots,z_{6}$ of colours in
$C^{\textup{left}}$ with each set
$z_{7},\dots,z_{13}$ of colours in
$C^{\textup{right}}$.

The technique of splitting regions and filtering out boundary colourings that
are guaranteed not to maximise $\mu_{1,2}(X)$ has a huge impact on
the running time of the program. On a fairly powerful home-PC as of
year 2006, it takes about two days to to obtain all 9440 values $\mu_{F_{f,m}}$.

\subsection{Constructing an $(\calA,\calF)$-set}
\label{sec:constructing-af-set}

\begin{figure}[tp]
\centering
(a)\hspace{-2mm}
\includegraphics[scale=\myscaling]{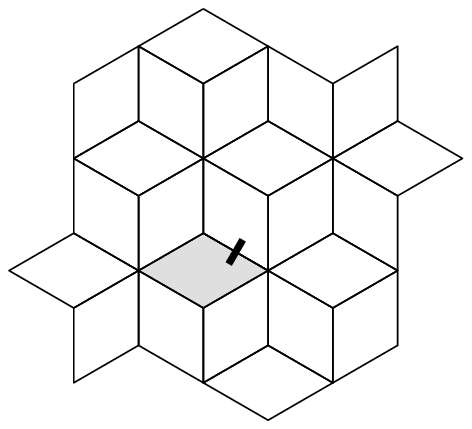}
(b)\hspace{-2mm}
\includegraphics[scale=\myscaling]{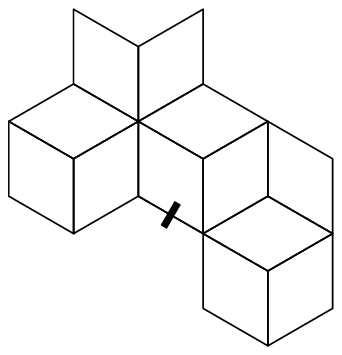}
(c)\hspace{-2mm}
\includegraphics[scale=\myscaling]{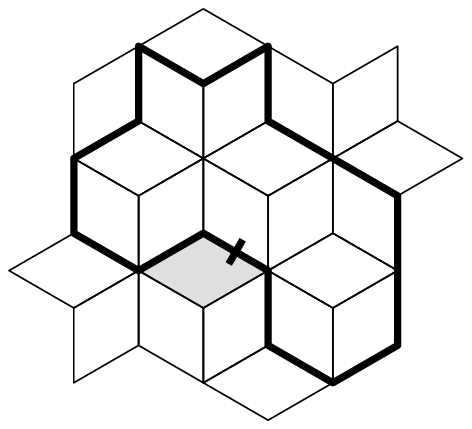}

\medskip{}

(d)\hspace{-2mm}\includegraphics[scale=\myscaling]{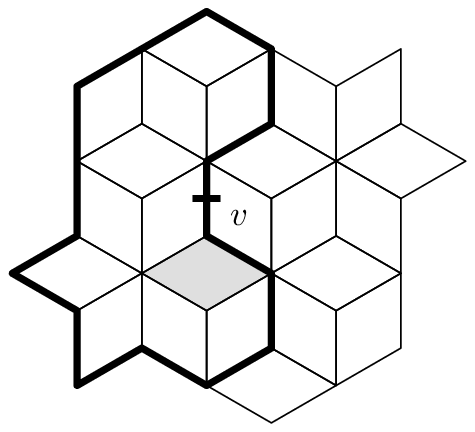}
(e)\hspace{-2mm}\includegraphics[scale=\myscaling]{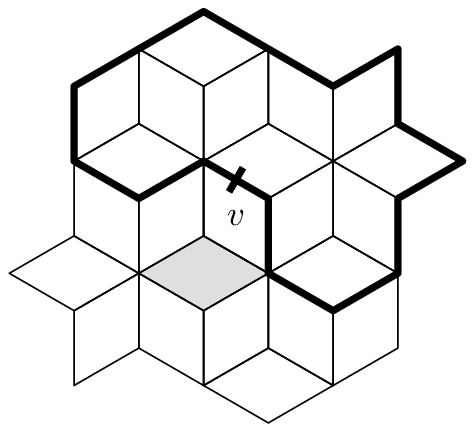}
(f)\hspace{-2mm}\includegraphics[scale=\myscaling]{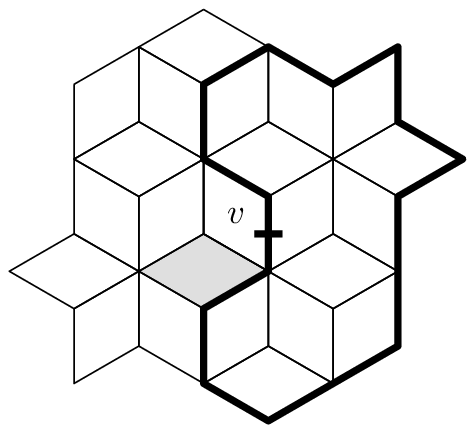}

\smallskip

\caption{\label{fig:kagome-big-and-all-moves}(a) The extended region
$\exR_{\textup{big}}$
(here all vertices are labelled ``in''). Note that the shaded
vertex is not a part of the region. (b) The extended region $\exR_{A}$
repeated. (c)--(f) Intersections of $\exR_{A}$ and $\exR_{\textup{big}}$.}

\end{figure}

We describe how to construct an $(\calA,\calF)$-set.
Let $\exR_{\textup{big}}$ be the extended region in
Figure~\ref{fig:kagome-big-and-all-moves}(a)
with some combination of labels ``in'' and ``out''
on the vertices.
From $\exR_{\textup{big}}$ we will derive 5-tuples that are added to a
set $\calS$.
By considering all possible combinations of labels ``in'' and ``out'' on the vertices
of $\exR_{\textup{big}}$,
we construct the $(\calA,\calF)$-set $\calS$.
We describe the process by first giving a concrete example.

Fix an ``in/out''-labelling of the vertices of the extended region $\exR_{\textup{big}}$.
Let $a \in \{1,\dots,342\}$ be the value such that
$\exR_{A_a}$ is an extended subregion of $\exR_{\textup{big}}$.
Note that
the extended regions $\exR_{A_i}$, $i \in \{1,\dots,342\}$,
are defined such that
there is exactly one value $a \in \{1,\dots,342\}$ for which this is true.
Figure~\ref{fig:kagome-big-and-all-moves}(b) shows the largest possible $\exR_{A_a}$
and
Figure~\ref{fig:kagome-big-and-all-moves}(c) shows the overlapping of
$\exR_{\textup{big}}$ and $\exR_{A_a}$.
We see from this figure that only some of the vertices of
$\exR_{\textup{big}}$ define $\exR_{A_a}$.
Similarly to how the extended region $\exR_{A_a}$ is obtained from
$\exR_{\textup{big}}$, let $a_1, a_2, a_3 \in \{1,\dots,342\}$ be the three
unique values such that $\exR_{A_{a_1}}$ is obtained from $\exR_{\textup{big}}$
by the overlapping in Figure~\ref{fig:kagome-big-and-all-moves}(d),
$\exR_{A_{a_2}}$ is obtained from $\exR_{\textup{big}}$
by the overlapping in Figure~\ref{fig:kagome-big-and-all-moves}(e),
and
$\exR_{A_{a_3}}$ is obtained from $\exR_{\textup{big}}$
by the overlapping in Figure~\ref{fig:kagome-big-and-all-moves}(f).
It is possible that
neighbours of vertex $v$ in Figure~\ref{fig:kagome-big-and-all-moves}(d)--(f)
are labelled ``out'',
meaning that some of the extended regions $\exR_{a_i}$ might not exist.
If this is the case we define $a_i = 0$
and $A_0 = \emptyset$.
For example, if the vertex to the left of vertex $v$ in
Figure~\ref{fig:kagome-big-and-all-moves}(d) is ``out'' then
$\exR_{a_1}$ cannot exist and hence $a_1 = 0$.

Suppose $X$ is an edge-boundary pair such that $R_X$ and
$\exR_{\textup{big}}$ are matching with respect to edge $e_X$.
Then $X \in A_a$.
For $i \in \{1,2,3\}$ and
any two distinct colours
$j, j' \in Q$ such that $p_{X}^{\textup{min}}(j,j') > 0$,
suppose $X_{i}(j,j')$ is the extended edge-boundary pair that is
constructed recursively in the tree $T_{X}$.
If $X_{i}(j,j') = \emptyset$ then $X_{i}(j,j') \in A_{a_i}$.
We will now be more precise about the sets of edge-boundary pairs and
incorporate the sets $M_1,\dots,M_4$.

\begin{figure}[tp]
\centering
\begin{tabular}{ccc}
(a)~~~\includegraphics[scale=\myscaling]{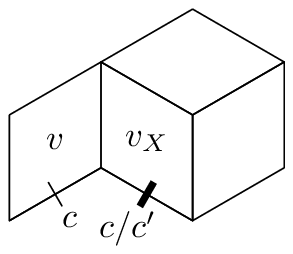}\phantom{(a)}
&
(e)~~~\includegraphics[scale=\myscaling]{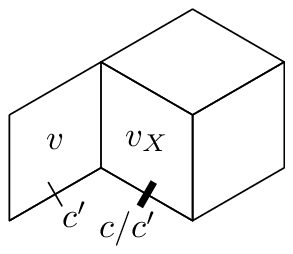}\phantom{(b)}
&
(i)~~~\includegraphics[scale=\myscaling]{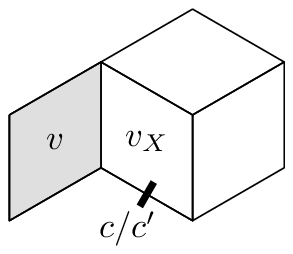}\phantom{(c)}\\
$(3,1,1)$, $(3,1,2)$, & $(4,1,1)$, $(4,1,2)$, & $(0,1,1)$, $(0,1,2)$,\\
$(3,2,1)$, $(3,2,2)$\phantom{,} & $(4,2,1)$, $(4,2,2)$\phantom{,} &
$(0,2,1)$, $(0,2,2)$\phantom{,}\\
\noalign{\vskip0.2cm}
(b)~~~\includegraphics[scale=\myscaling]{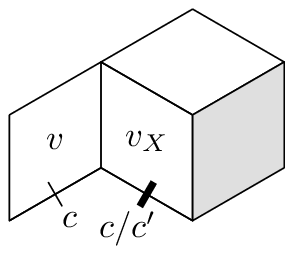}
&
(f)~~~\includegraphics[scale=\myscaling]{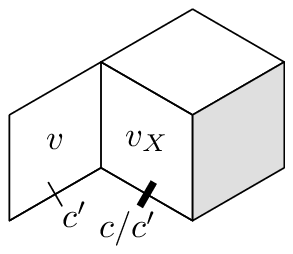}
&
(j)~~~\includegraphics[scale=\myscaling]{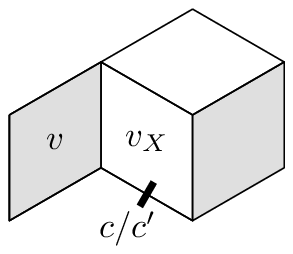}\\
\hspace{6mm}$(3,4,0)$\hspace{6mm} & \hspace{6mm}$(4,4,0)$\hspace{6mm} &
\hspace{6mm}$(0,4,0)$\hspace{6mm}\\
\noalign{\vskip0.2cm}
(c)~~~\includegraphics[scale=\myscaling]{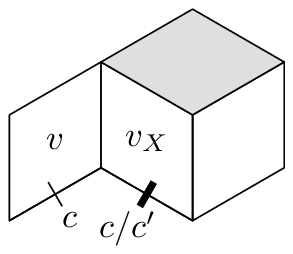}
&
(g)~~~\includegraphics[scale=\myscaling]{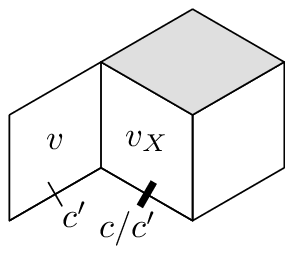}
&
(k)~~~\includegraphics[scale=\myscaling]{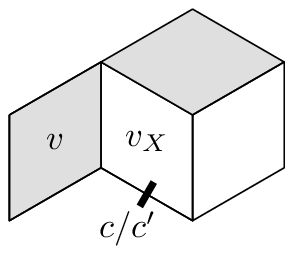}\\
$(3,0,4)$ & $(4,0,4)$ & $(0,0,4)$\\
\noalign{\vskip0.2cm}
(d)~~~\includegraphics[scale=\myscaling]{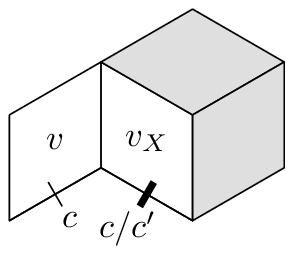}
&
(h)~~~\includegraphics[scale=\myscaling]{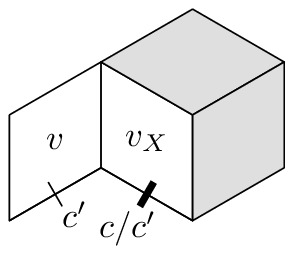}
&
(l)~~~\includegraphics[scale=\myscaling]{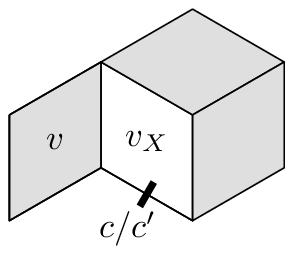}\\
$(3,0,0)$ & $(4,0,0)$ & $(0,0,0)$
\end{tabular}

\caption[Cases covering which sets $M_{1},\dots,M_{4}$ edge-boundary pairs
belong to]{\label{fig:kagome-twelve-cases}
Twelve cases which cover all possible
combinations of the sets $M_{1},\dots,M_{4}$ to which the recursively
constructed edge-boundary pairs $X_{1}(c_{1},c_{2})$, $X_{2}(c_{1},c_{2})$
and $X_{3}(c_{1},c_{2})$ belong.
If $X \in M_1$ then (a)--(d) apply.
If $X \in M_2$ then (e)--(h) apply.
If $X \in M_3 \cup M_4$ then (i)--(l) apply.}
\end{figure}

Suppose without loss of generality that
$B_X(e_X) = c$ and $B'_X(e_X) = c'$, and
$\mu_{c,c'}(X) \geq \mu_{c',c}(X)$
for some $c, c' \in Q$.
Suppose the extended region in
Figure~\ref{fig:kagome-twelve-cases}(a)
is an extended subregion of $\exR_{\textup{big}}$.
Suppose that the colour of the edge between $w_X$ and $v$ in
Figure~\ref{fig:kagome-twelve-cases}(a)
has colour $c$ in $B_X$ and $B'_X$.
Then $X \in M_1$ and hence $X \in A_{a,1}$.
From
Figure~\ref{fig:kagome-twelve-cases}(a)
we see that the extended region $\exR_{M_{(3,4)}}$
in Figure~\ref{fig:kagome-four-sets}(b)
is an
extended subregion of $\exR_{A_{a_1}}$.
Hence $X_{1}(j,j')$ belongs to $M_3$ or $M_4$ (or both).
The crucial observation here is that
$p_{X}^{\textup{min}}(j,j') > 0$ if and only if $j' = c$.
This follows from the fact that
$\mu_{c,c'}(X) \geq \mu_{c',c}(X)$
and hence there is a discrepancy at $v_X$ only when
the colour $c$ is drawn from $\pi_{B'_X}$ in the coupling
$\Psi_X^\textup{min}$.
We therefore conclude that $X_{1}(j,j') \in M_3$. Thus,
$X_{1}(j,j') \in A_{a_1,3}$.
For $X_{2}(j,j')$ and $X_{3}(j,j')$
we see in Figure~\ref{fig:kagome-twelve-cases}(a)
that these edge-boundary pairs belong to either $M_1$ or $M_2$.
However, we are unable to tell exactly to which of the two sets
these edge-boundary pairs belong.
We therefore assume that any combination of the two sets is possible.
The 3-tuples listed in
Figure~\ref{fig:kagome-twelve-cases}(a)
indicate to which possible sets $M_1,\dots,M_4$ the edge boundary pairs
$X_{1}(j,j')$, $X_{2}(j,j')$ and $X_{3}(j,j')$ belong.
That is, a 3-tuple $(m_1,m_2,m_3)$ means that
$X_{1}(j,j') \in A_{a_1,m_1}$,
$X_{2}(j,j') \in A_{a_2,m_2}$ and
$X_{3}(j,j') \in A_{a_3,m_3}$.

Let $\calF' \subseteq \calF$ be the set of edge-boundary pairs $F_{i,1}$
such that $\exR_{F_i}$ is
an extended subregion of $\exR_{\textup{big}}$
and $i \in \{1,\dots,4720\}$.
Then $X \in F_{i,1}$ for every $F_{i,1} \in \calF'$.
Remember that we have assumed above that $X \in M_1$.
Let $f$ be the value such that
$F_{f,1} \in \calF'$ is the set that minimises $\mu_{F_{i,1}}$ over all
$F_{i,1} \in \calF'$. If the minimiser is not unique, let $f$ be the smallest $i$
among the minimisers.
Now, for each 3-tuple $(m_1,m_2,m_3)$ in
Figure~\ref{fig:kagome-twelve-cases}(a)
we add the following 5-tuple to the set $\calS$:
$(A_{a,1}, F_{f,1},\nolinebreak[3] A_{a_1,m_1},\nolinebreak[3] A_{a_2,m_2}, A_{a_2,m_2})$.

Summing it all up, we construct the set $\calS$ as follows.
First take an extended region $\exR_{\textup{big}}$.
From $\exR_{\textup{big}}$ we uniquely derive the sets
$A_a$, $A_{a_1}$, $A_{a_2}$ and $A_{a_3}$.
If $\exR_{M_{(1,2)}}$ is an extended subregion of $\exR_{\textup{big}}$
then we consider two values of $m$: $m=1$ and $m=2$.
If $\exR_{M_{(3,4)}}$ is an extended subregion of $\exR_{\textup{big}}$
then we also consider two values of $m$: $m=3$ and $m=4$.
Now suppose $X \in A_{a,m}$.
The twelve cases in
Figure~\ref{fig:kagome-twelve-cases}
cover all possible combinations of the sets $M_1,\dots,M_4$ to which
the recursively constructed edge-boundary pairs
$X_{1}(j,j')$, $X_{2}(j,j')$ and $X_{3}(j,j')$ belong.
More precisely,
if $m = 1$ then
Figure~\ref{fig:kagome-twelve-cases}(a)--(d) apply.
if $m = 2$ then
Figure~\ref{fig:kagome-twelve-cases}(e)--(h) apply.
if $m = 3$ or $m = 4$ then
Figure~\ref{fig:kagome-twelve-cases}(i)--(l) apply.
From $\exR_{\textup{big}}$ and the value of $m$, we uniquely derive the set
$F_{f,m}$ to which $X$ belongs.
For each 3-tuple $(m_1,m_2,m_3)$ in the relevant case in
Figure~\ref{fig:kagome-twelve-cases},
we add the following 5-tuple to the set $\calS$:
$(A_{a,m}, F_{f,m},\nolinebreak[3] A_{a_1,m_1},\nolinebreak[3] A_{a_2,m_2}, A_{a_2,m_2})$.
If the value of $m_i$ in a 3-tuple is~0 then $A_{a_i,m_i} = \emptyset$.
By considering every possible extended region $\exR_{\textup{big}}$ and
every possible value of $m$ (two values per region $\exR_{\textup{big}}$),
we construct a set $\calS$ that is an $(\calA,\calF)$-set.


\end{document}